\documentclass[twocolumn,fleqn]{aastex61}
\submitjournal{ApJ}

\usepackage{amssymb}
\usepackage[fleqn]{amsmath}
\usepackage[varg]{txfonts}
\usepackage{natbib}
\usepackage{url}             % For breaking URLs easily through lines
\usepackage{graphicx}
\usepackage{float}
\newcommand{\beq}{\begin{equation}}
\newcommand{\eeq}{\end{equation}}
\newcommand{\beqa}{\begin{eqnarray}}
\newcommand{\eeqa}{\end{eqnarray}}
\newcommand{\beqann}{\begin{eqnarray*}}
\newcommand{\eeqann}{\end{eqnarray*}}
\bibpunct{(}{)}{;}{a}{}{,} % to follow the A&A style

\usepackage{textcomp}
\usepackage{wrapfig}

\shorttitle{Simulations of two-fluid spicules}
\shortauthors{Ku\'zma et al.}
\begin{document}

%\titlerunning{Simulations of 2-fluid spicules}
\title{Two-fluid numerical simulations of solar spicules% \\
%in the framework of a 2-fluid model
}

%\offprints{B.~Ku\'zma}

%\author{B.~Ku\'zma\inst{\ref{inst1}}
%\and K.~Murawski\inst{\ref{inst1}}
%\and P.~Kashyap\inst{\ref{inst1}}
%\and D.~W\'ojcik\inst{\ref{inst1}}
%}

%\offprints{B.~Ku\'zma}

%\institute{$^1$Group of Astrophysics, University of Maria Curie-Sk{\l}odowska, ul. Radziszewskiego 10, 20-031 Lublin, Poland \label{inst1}
%}
\correspondingauthor{B{\l}a\.zej Ku\'zma}
\email{blazejkuzma1@gmail.com}

%\author[0000-0002-0786-7307]{Greg J. Schwarz}I SHOULD ADD ORCID
\author{B{\l}a\.zej Ku\'zma}
\affil{Group of Astrophysics, University of Maria Curie-Sk{\l}odowska, ul. Radziszewskiego 10, 20-031 Lublin, Poland}

\author{Kris Murawski}
\affiliation{Group of Astrophysics, University of Maria Curie-Sk{\l}odowska, ul. Radziszewskiego 10, 20-031 Lublin, Poland}

\author{Pradeep Kayshap}
\affiliation{Group of Astrophysics, University of Maria Curie-Sk{\l}odowska, ul. Radziszewskiego 10, 20-031 Lublin, Poland}

\author{Darek W\'ojcik}
\affiliation{Group of Astrophysics, University of Maria Curie-Sk{\l}odowska, ul. Radziszewskiego 10, 20-031 Lublin, Poland}

\author{Abhishek Kumar Srivastava}
\affiliation{Department of Physics, Indian Institute of Technology (BHU), Varanasi-221005, India}

\author{Bhola N. Dwivedi}
\affiliation{Department of Physics, Indian Institute of Technology (BHU), Varanasi-221005, India}

%PLEASE DO NOT DELETE
%\abstract
%{}
%{
%We aim to study the formation and evolution of solar spicules using 2-fluid numerical simulations of the solar atmosphere. 
%}
%{
%With the use of JOANNA code, we numerically solve 2-fluid (for ions and neutrals) equations in 2D Cartesian geometry. We follow the evolution of %spicules triggered by pulses in vertical velocity component launched from the upper chromosphere. 
%}
%{
%Our numerical results reveal that the velocity pulses are steepened into a shock that propagates upward into the corona. The chromospheric cold %and dense plasma follows the shock and rises into the corona with the mean speed of 20-25 km s$^{-1}$. The formed spicule shows the upflow/%downfall of plasma within its total life-time of around 3 minutes, and it follows the typical characteristics of classical spicule. The spicule %detected in neutrals is of higher height and rise time than the spicule seen in ions, which is the result of high pressure gradient in neutrals %compared to the ions. The formed spicule consists of a dense cold core, where ions and neutrals remain in balance, and its upper part becomes %highly ionized. We found that the center and bottom of the spicule-core is surrounded by neutrals-dominated halo. 
%}
%{
%In the framework of the 2-fluid, 2D model we devised, we investigated evolution and described structure of solar spicules. 
%}
%%PLEASE DO NOT DELETE

%\keywords{Sun: activity - Sun: corona - Sun: transition region - magnetohydrodynamics (MHD) - methods: numerical}
\begin{abstract}

We aim to study the formation and evolution of solar spicules by means of numerical simulations of the solar atmosphere. With the use of newly developed JOANNA code, we numerically solve two-fluid (for ions + electrons and neutrals) equations in 2D Cartesian geometry. We follow the evolution of a spicule triggered by the 
%a pulses in pressure and vertical components of ion and neutral velocity 
time-dependent signal in ion and neutral components of gas pressure 
launched in the upper chromosphere. We use the potential magnetic field, which evolves self-consistently, but mainly plays a passive role in the dynamics. Our numerical results reveal that the signal is steepened into a shock that propagates upward into the corona. The chromospheric cold and dense plasma lags behind this shock and rises into the corona with a mean speed of 20-25 km s$^{-1}$. The formed spicule exhibits the upflow/downfall of plasma during its total lifetime of around 3-4 minutes, and it follows the typical characteristics of a classical spicule, which is modeled by magnetohydrodynamics. The simulated spicule consists of a dense and cold core that is dominated by neutrals. The general dynamics of ion and neutral spicules are very similar to each other. Minor differences in those dynamics result in different widths of both spicules with increasing rarefaction of the ion spicule in time. 

%in the spicule detected in ions is thiner and becomes highly rarefied in time. 
%The spicule detected in neutrals is of higher height and rise time than the spicule seen in ions, which is the result of high pressure gradient in neutrals compared to the ions. 
%\textbf{The formed spicule consists of a dense cold core, where ions and neutrals remain in balance, and its upper part becomes highly ionized. We found that the center and bottom of the spicule-core is surrounded by neutrals-dominated halo.} 

\end{abstract}


%\titlerunning{Simulations of 2-fluid spicules}
%\authorrunning{B.~Ku\'zma et al.}

%\maketitle
% -----------------------------------------------------------------------------------------
\section{Introduction}
% -----------------------------------------------------------------------------------------
Spicules are thin jet-like structures, which dominate in the lower layers of the solar corona. They are best seen at the solar limb in strong chromospheric and transition-region (TR) spectral lines such as H$\alpha$, Ca~{\sc ii} H $\&$ K, Mg~{\sc ii} H $\&$ K, C~{\sc ii} and Si~{\sc iv} lines (e.g., \citealt{Roberts1945,Matsuno1988,Nisikawa1988, Suematsu1995,DePon2007, Suematsu2008,Sterling2010,Mad2011,Per2012,Tsir2012,Skog2014,Skog2015,DePon2014,Rouppe2015,Per2016,Beck2016}). 
Spicule observations have been carried out for about 130 years since they were first reported in 1877 (\citealt{Secchi1877}). Therefore, a huge amount of observational literature about spicules is available, which is dedicated to understanding %such aspects of spicules as 
their basic properties (mass density, temperature, velocity, and magnetic field), initiation mechanisms, waves, and oscillations. These crucial aspects are very well documented in various review papers, i.e., basic properties by \cite{Beckers1968,Beckers1972} and \cite{Suematsu1998}, initiation mechanisms by \cite{Sterling2000}, and oscillations and waves by \cite{Zaq2009}.  

High-resolution observational data leads to continuous improvement of our knowledge about the spicules. Based on their properties, \cite{DePon2007} classified the spicules into two different categories (i.e., Types I and II). Type I spicules exhibit a slower velocity (15-40 km s$^{-1}$) and longer lifetime (3-10 minutes) compared to the velocity (30-110 km s$^{-1}$) and lifetime (50-150 s) of Type II spicules. Type I spicules reveal the rise and fall of plasma during its total life, while type II spicules fade from the view and the downfall of plasma is not visible (e.g., \citealt{DePon2007,Per2012}). \cite{Sterling1984} have proposed that the heating rate is sufficient to heat the spicules to temperatures at which the hydrogen is fully ionized. Therefore, H-$\alpha$ spicules (cool temperature) may evolve into EUV spicules (hot temperature). Later, this thermal evolution (i.e., fading of the spicules from cool filters and appears in the hot filters) of the spicules was investigated in various other works as well (e.g., \citealt{DePon2009,McIntosh2009,McIntosh2010,Tian2011}). Recently, \cite{DePon2014} have reported the traces of spicules in the hot temperature filter after fading from the cool temperature filter.  

The origin and excitation mechanisms of spicules are one of the most crucial issues of solar physics, which has been continuously investigated using observations as well as numerical simulations. Pulses in velocity or gas pressure, Alfv{\'e}n waves, and p-modes are the three broad categories of drivers that may be responsible for the formation of spicules. Main physical process within the velocity/gas pressure pulse model is the formation of a shock front, which results in a generation of the spicule in the lower corona. Such a rebound shock model was developed by \cite{Hollweg1982} who used a gas pressure pulse in one-dimensional (1D) MHD equations for the formation of a spicule. In another approach, \cite{Suematsu1982} performed a numerical experiment using a velocity pulse instead of a gas pressure pulse. Later, this rebound shock model was improved by including radiation and heat conduction in the model (e.g., \citealt{Sterling1990,Cheng1992,Sterling1993,Heggland2007,Kuzma2017}). Recently, \cite{Sterling2016} have reported that the microfilament-eruptions can be a potential candidate for the generation of spicules. \cite{Murawski2010} performed 2D numerical simulations with a velocity pulse to produce the spicule. Multistructural, bidirectional flows and 3-5 minute periodicity in the occurrence of spicules were successfully reproduced in that numerical experiment. This model was extended into its 2D counterpart with the inclusion of the nonadiabatic effects (\citealt{Kuzma2017}). A general conclusion drawn from the performed simulations was that the shocks play an important role in the dynamics and thermodynamics of impulsively generated spicules and the other nonadiabatic terms (i.e., thermal conduction and radiative cooling terms) produce minor effects on the dynamics/thermodynamics of the spicule.  

In the second category, it was proposed that the nonlinear coupling between Alfv{\'e}n waves and slow magnetoacoustic shocks can lift up the TR, which produces the spicule (\citealt{Hollweg1982a}). This idea for the generation of spicules was further investigated by \citet{Cranmer2015,Cranmer2016}. They have found that the magnetohydrodynamic turbulence is a potential driver for spicules. It was also reported that the random nonlinear Alfv{\'e}n pulses may reproduce the spicules (\citealt{Kudoh1999}). More interestingly, the damping of Alfv{\'e}n waves due to ion-neutral collisions can be an efficient mechanism for the formation of spicules in the solar atmosphere (\citealt{Hae1992,James2003}). 

In the third category, \cite{DePon2004} showed that p-modes are the potential candidates for triggering spicules in the solar atmosphere, which was further investigated in various other works (e.g., \citealt{Hansteen2006, DePon2007a}). 

The above mentioned main three categories (i.e., gas pressure or velocity pulse, Alfv{\'e}n waves and p-modes) are not the only mechanisms for the formation of spicules. Indeed, there are various other numerous proposed physical processes responsible for the formation of spicules. Among others, the compression of the plasma sheet by the magnetic field (\citealt{Hollweg1972}), Joule heating in the current sheet (\citealt{Hirayama1992}), thermal conduction from the corona (\citealt{Kopp1968,Moore1972}) and buffeting of anchored magnetic flux by granulation can lead to spicules (\citealt{Roberts1979}). These and various other mechanisms are discussed by \cite{Sterling2000}.

Although spicules were discovered a long time ago (\citealt{Secchi1877}), they are still not fully understood. Therefore, they remain subject 
%a mystery for the solar/space physics research 
%On the basis of observations as well as advanced computing facilities, a significant progress has been achieved in the understanding of drives of spicules. As the present status of our knowledge on the spicule's drivers is very far away from the satisfactory level, 
%and therefore it remains 
of further investigation. 
Advanced numerical simulations are an important means to reveal the nature of spicules. However, it is always difficult to implement the real solar conditions in the numerical experiment. 
Because low layers of the solar atmosphere contain large fraction of neutrals (e.g., \citealt{Zaqarashvili2011}), two-fluid plasma (i.e., ionized and neutral fluids) approach 
in the numerical experiment is more suitable than magnetohydrodynamics (MHD) to study the evolution of spicules in the solar atmosphere. This approach is used for the first time in the present numerical simulations of solar spicules.
%, which is one step further towards the real solar conditions unlike the previous numerical simulations which were based on single fluid approach. 
Specifically, we perform 2D numerical simulations of two-fluid equations to produce the spicule using pulses launched initially in vertical components of ion and neutral velocities. The adopted potential magnetic field configuration plays essentially a passive role in the present simulations. %The simulated spicule has cold core with the balance between ions and neutrals, however, the upper part of spicule becomes highly ionized. 
The paper is organized as follows. In Section~\ref{sec:atm_model}, we describe the physical model of the solar atmosphere. Numerical simulations %of MHD equations 
are presented in Section~\ref{sec:num_sim_MHD}. Summary and conclusions are outlined in the last section.   
% -----------------------------------------------------------------------------------------
\section{Physical Model of the Solar Atmosphere}\label{sec:atm_model}
\subsection{Two-fluid Equations}\label{sec:equ_model}
% -----------------------------------------------------------------------------------------

% -----------------------------------------------------------------------------------------
We consider a gravitationally stratified and magnetically confined plasma that consists of two components: ionized fluid (ions + electrons) and neutral fluid (neutrals). This model is governed by the following set of equations (\citealt{Sakai2008}): %(\citealt{Zaqarashvili2011}):

\begin{equation}
\label{eq:MHD_rho}
\begin{aligned}
&{{\partial \varrho_{i}}\over {\partial t}}+\nabla \cdot (\varrho_{i}{{\bf V} _{i}})=-\varrho_{i}(\alpha_{r}\varrho_{i}-\alpha_{i}\varrho_{n})\, ,
\end{aligned}
\end{equation}
\begin{equation}
\label{eq:MHD_V}
\begin{aligned}
&{{\partial \varrho_{n}}\over {\partial t}}+\nabla \cdot (\varrho_{n}{{\bf V} _{n}})=\varrho_{i}(\alpha_{r}\varrho_{i}-\alpha_{i}\varrho_{n})\, ,\\
\end{aligned}
\end{equation}
\begin{equation}
\label{eq:MHD_V2}
\begin{split}
\begin{aligned}
&\varrho_{i}\left(\frac{\partial{{\bf V} _{i}}}{\partial t} + ({{\bf V} _{i}}\cdot \nabla){{\bf V} _{i}}\right)=-\nabla p_{i}+\frac{1}{\mu} (\nabla \times {\bf B}) \times {\bf B}+\varrho_{i}{\bf g}  \\
&-\alpha_{c}\varrho_{i}\varrho_{n}({{\bf V}_{i}}-{{\bf V}_{n}}) -\varrho_{i}(\alpha_{r}\varrho_{i}{\bf V}_{i}-\alpha_{i}\varrho_{n}{\bf V}_{n})\, , \\
\end{aligned}
\end{split}
\end{equation}
\begin{equation}
\label{eq:MHD_p}
\begin{split}
\begin{aligned}
&\varrho_{n}\left(\frac{\partial{{\bf V} _{n}}}{\partial t} + ({{\bf V} _{n}}\cdot \nabla){{\bf V} _{i}}\right)=-\nabla p_{n} +\varrho_{n} {\bf g} \\
&+\alpha_{c}\varrho_{i}\varrho_{n}({{\bf V}_{i}}-{{\bf V}_{n}}) +\varrho_{i}(\alpha_{r}\varrho_{i}{\bf V}_{i}-\alpha_{i}\varrho_{n}{\bf V}_{n})\, , \\
\end{aligned}
\end{split}
\end{equation}
\begin{equation}
\label{eq:MHD_p2}
\begin{split}
\begin{aligned}
&\frac{\partial p_{i}}{\partial t} + {{\bf V}_{i}} \cdot \nabla p_{i} +\gamma p_{i}\nabla \cdot {{\bf V}_{i}} \\
&= (\gamma-1)\alpha_{c}\varrho_{i}\varrho_{n}({{\bf V} _{i}}-{{\bf V} _{n}})\cdot{{\bf V}_{i}}\, , 
%+ \\ 
%(\gamma -1)\alpha_{en}({{\bf V} _{e}}-{{\bf V} _{n}})\cdot{{\bf V}_{e}}\, , 
\end{aligned}
\end{split}
\end{equation}
\begin{equation}
\label{eq:MHD_B}
\begin{aligned}
\begin{split}
%\frac{\partial p_{n}}{\partial t} + {{\bf V}_{n}} \cdot \nabla p_{n} +\gamma p_{n}\nabla \cdot {{\bf V}_{n}}
%= (\gamma-1)\alpha_{in}({{\bf V} _{i}}-{{\bf V} _{n}})\cdot{{\bf V}_{n}} - \\ 
%(\gamma -1)\alpha_{en}({{\bf V} _{e}}-{{\bf V} _{n}})\cdot{{\bf V}_{n}}\, , 
&\frac{\partial p_{n}}{\partial t} + {{\bf V}_{n}} \cdot \nabla p_{n} +\gamma p_{n}\nabla \cdot {{\bf V}_{n}} \\
& = (\gamma-1)\alpha_{c}\varrho_{i}\varrho_{n}({{\bf V} _{i}}-{{\bf V} _{n}})\cdot{{\bf V}_{n}} \, ,
\end{split}
\end{aligned}
%\label{eq:MHD_B}
\end{equation}
\begin{equation}
\begin{aligned}
&\frac{\partial {\bf B}}{\partial t} = \nabla \times ({\bf V} \times {\bf B})\, , \, \nabla \cdot {\bf B}=0 \, ,
%{\bf j}=\frac{1}{\mu} \nabla \times {\bf B} \, ,\\
\end{aligned}
\label{eq:MHD_B2}
\end{equation}
% -----------------------------------------------------------------------------------------
%where ${n}$ is the number of particles, $m$ the mass of particles, $p$ the gas pressure, 
where $\varrho_{i,n}$ is the mass density, $p_{i,n}$ the gas pressure, 
%${\bf j}$ current density, 
${\bf V}_{i,n}$ represents the plasma velocity, ${\bf B}$ is the magnetic field, $\alpha_{i}$ is a coefficient of ionization, $\alpha_{r}$ coefficient of recombination, and $\alpha_{c}$ coefficient of collisions between particles, subscripts %$e$, 
$i$ and $n$ correspond, respectively, to 
%electrons, 
ions and neutrals, 
%in $\varrho_{i,n}$, $p_{i,n}$ and ${\bf V}_{i,n}$, 
$\gamma=5/3$ is the adiabatic index, and ${\bf g}=(0,-g,0)$ is the gravitational acceleration. The value of $g$ is equal to $274$ m s$^{-2}$. For the sake of simplicity, all electron-components are neglected due to the small mass of electrons in relation to ions and neutrals. %as well as ionization and recombination terms are not taken into account in Eqs. (1) \& (2). 

%The symbol $k_{\rm B}$ denotes the Boltzmann constant, $\gamma=5/3$ is the adiabatic index, $m$ is the particle mass that is specified by a mean molecular weight of $0.6$, and ${\bf g}=(0,-g,0)$ is the gravitational acceleration. The value of $g$ is equal to $274$ m s$^{-2}$.

% -----------------------------------------------------------------------------------------
\subsection{Equilibrium Solar Atmosphere}\label{sec:equil}
% -----------------------------------------------------------------------------------------

% -----------------------------------------------------------------------------------------
\begin{figure}[!hb]
        \begin{center}
        \hspace{0.5cm}
        \mbox{
                \includegraphics[scale=0.45, angle=0]{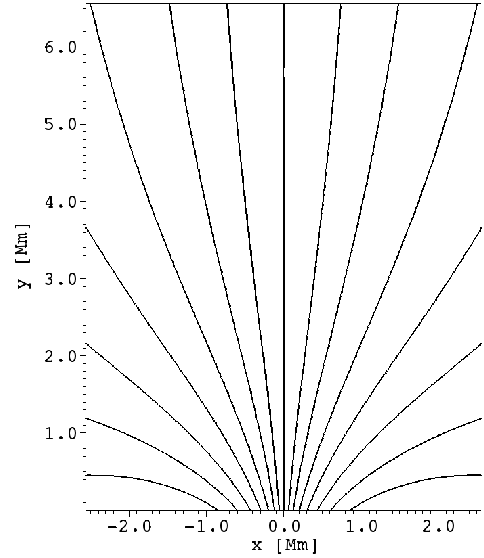}
                }
\vspace{-0.5cm}
                \caption{\small Magnetic field lines %and the average temperature $T$ (color map), expressed in Kelvins, 
                at the plasma equilibrium.
                        }
                \label{fig:2}
        \end{center}
\end{figure}
% -----------------------------------------------------------------------------------------

\begin{figure}[!ht]
        \begin{center}
        \hspace{-0.7cm}
                \includegraphics[scale=0.5, angle=0]{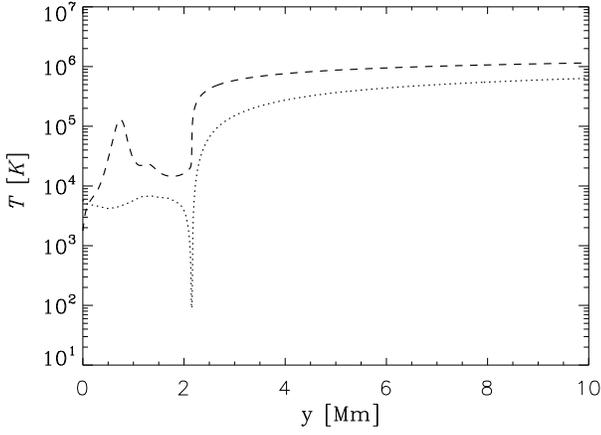}
                \caption{\small Vertical profiles of hydrostatic temperature of ions (dashed line) and neutrals (dotted line) vs. height $y$.
                        }
                \label{fig:1}
        \end{center}
\end{figure}

In a static solar atmosphere, all plasma quantities are time-invariant, which means that $\partial f / \partial t = 0$, where $f$ denotes an equilibrium plasma quantity. Then, from Equations~(\ref{eq:MHD_rho})-(\ref{eq:MHD_B2}), it follows that for still ions $({\bf V}_{i} = {\bf 0})$ and neutrals $({\bf V}_{n} = {\bf 0})$ the Lorentz force (if appropriate) must be balanced by the gravity force and the gas pressure gradient,
% -----------------------------------------------------------------------------------------
\begin{equation}
\begin{aligned}
\label{eq:B}
&\frac{1}{\mu}(\nabla \times{{\bf B}})\times{{\bf B}} - \nabla p_{i,n}+ \varrho_{i,n} {\bf g}
= {\bf 0} \, . 
%- \nabla p_{n} + \varrho _{n} {\bf g} = 0\, .
\end{aligned}
\end{equation}

% -----------------------------------------------------------------------------------------

% -----------------------------------------------------------------------------------------

% -----------------------------------------------------------------------------------------

% -----------------------------------------------------------------------------------------
%This model of the solar atmosphere corresponds to a quiet Sun.

% -----------------------------------------------------------------------------------------
\subsubsection{Current-Free Magnetic Field and the Hydrostatic Atmosphere}
% -----------------------------------------------------------------------------------------

A hydrostatic atmosphere for ions corresponds to the force-free $((\nabla \times {\bf B})\times {\bf B}={\bf 0})$ magnetic field with the solenoidal condition $(\nabla \cdot {\bf B}=0)$ being satisfied. We additionally assume a current-free $(\nabla \times {\bf B}={\bf 0})$ magnetic field whose horizontal $B_{x}$, vertical $B_{y}$, and transversal $B_{z}$ components are given as (\citealt{Konkol2012}) 
% -----------------------------------------------------------------------------------------
\begin{equation}
\begin{split}
\begin{aligned}
&B_{x}(x,y)=\frac{-2Sx(y-a)}{(x^2+(y-a)^2)^{2}} \, , \hspace{3mm} \\
&B_{y}(x,y)=\frac{S(x^2-(y-a)^2)}{(x^2+(y-a)^2)^{2}}+B_{\rm v} \, , \hspace{3mm} \\
&B_{z}(x,y)=0 \, ,
\end{aligned}
\end{split}
\end{equation}
% -----------------------------------------------------------------------------------------
where $B_{\rm v}$ is a straight vertical magnetic field component, $a$ and $S$ are free parameters corresponding to the vertical location of the singularity in the magnetic field and the magnetic field strength, respectively. We set  $B_{\rm v}=6$ Gs, $a=-1.5$ Mm and $S$ in such a way that at the reference point $(x_{r}= 0, y_{r} = 10)$ Mm the magnitude of magnetic field $B=8$ Gs. The corresponding magnetic field lines are displayed in Figure~\ref{fig:2}. We
note that the magnetic field lines become less divergent with height.

\begin{figure*}[!ht]
        \begin{center}
        \mbox{
        \hspace*{-1.5cm}
                \includegraphics[scale=0.4, angle=0]{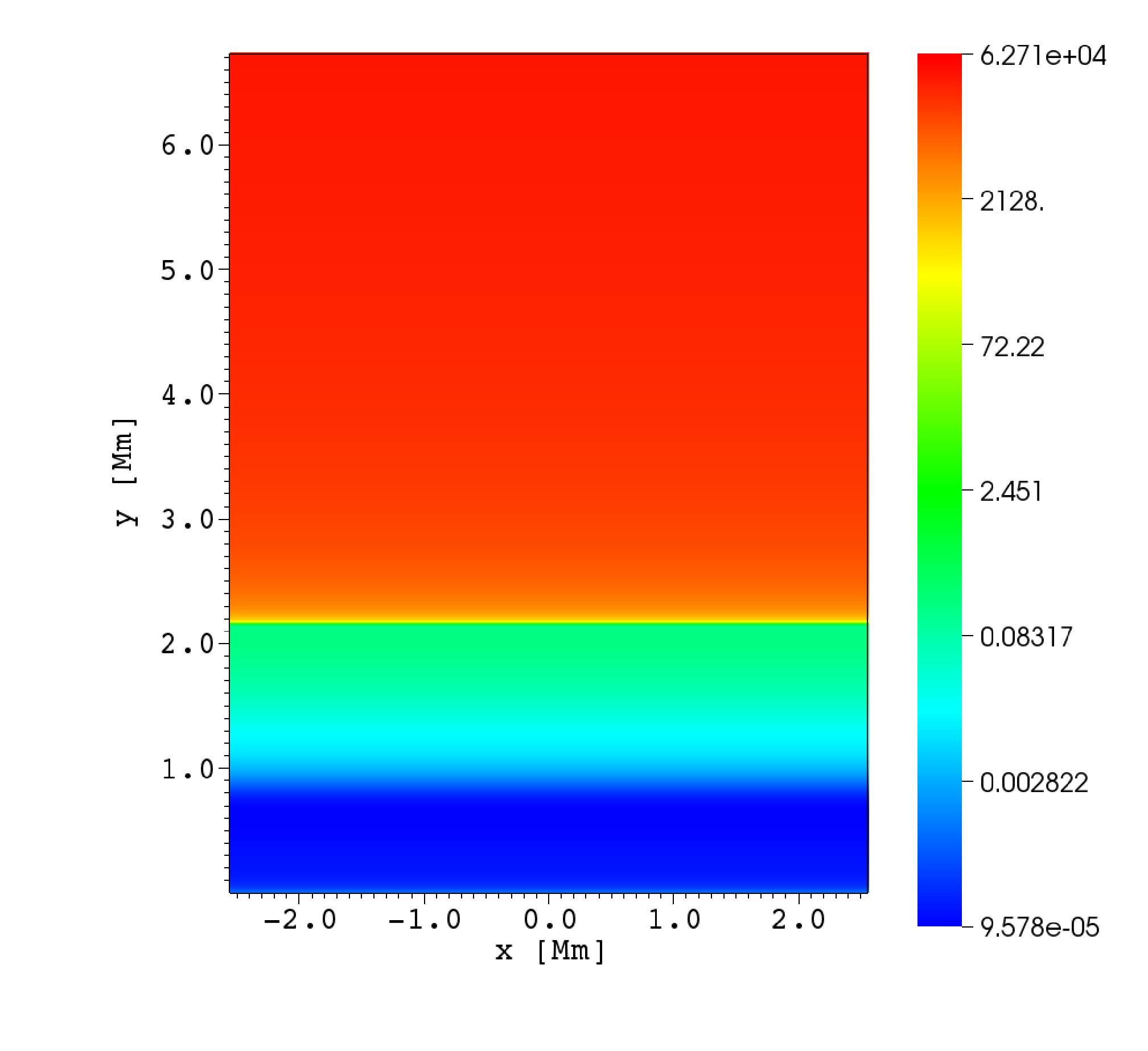}\hspace*{-0.0cm}
                \includegraphics[scale=0.4, angle=0]{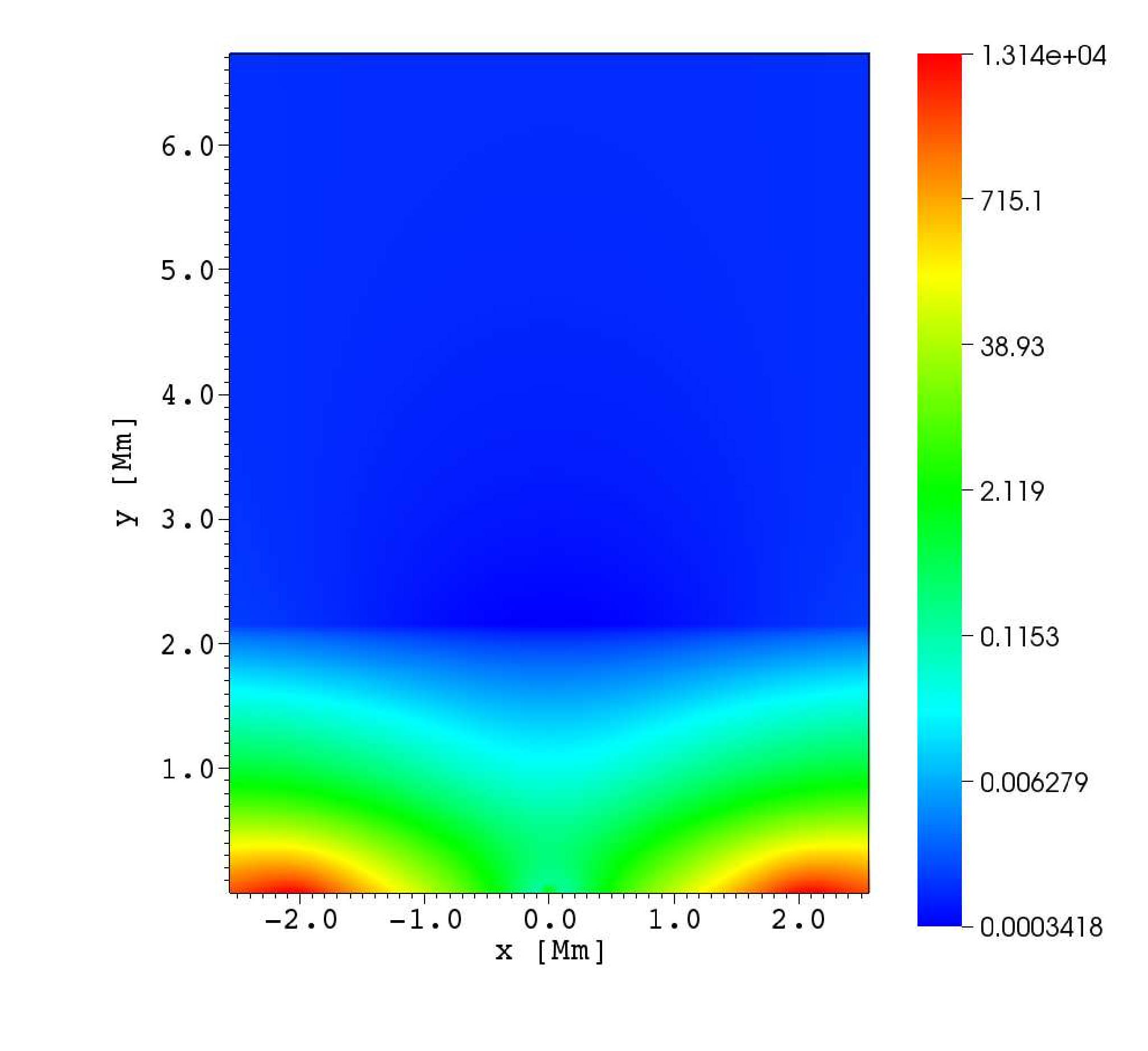}
                }
%\vspace{1.4cm}
                \caption{\small Ratio of the mass density of ions to the mass density of neutrals (left) and plasma beta (right) at the plasma equilibrium.}
                \label{fig:3}
        \end{center}
\end{figure*}
% -----------------------------------------------------------------------------------------

For a force-free magnetic field it follows from  Eq.~(\ref{eq:B}) that at the equilibrium the gas pressure gradients have to be  balanced by the gravity force,
% -----------------------------------------------------------------------------------------.
\begin{equation}
\begin{aligned}
&\nabla p_{i,n} = \varrho _{i,n} {\bf g} \, .
\label{eq:B568}
\end{aligned}
\end{equation}
% -----------------------------------------------------------------------------------------
%The subscript h corresponds to a hydrostatic quantity. 
With the use of the ideal gas law and the vertical, $y$-component of Eq.~(\ref{eq:B568}), we express the hydrostatic gas pressures and mass densities of ions and neutrals as
% -----------------------------------------------------------------------------------------
\begin{equation}
\label{eq:pres}
\begin{split}
\begin{aligned}
&p_{i,n}(y)=p_{0 \, i,n}~{\rm exp}\left( -
\int_{y_{\rm r}}^{y}\frac{dy^{'}}{\Lambda _{i,n} (y^{'})} \right)\, , \\
&\varrho_{i,n} (y)=\frac{p_{i,n} (y)}{g \Lambda _{i,n} (y)}\, ,
\end{aligned}
\end{split}
\end{equation}
% -----------------------------------------------------------------------------------------
where
% -----------------------------------------------------------------------------------------
\begin{equation}\label{eq:lambda}
\Lambda_{i,n}(y) = \frac{k_{\rm B} T_{i,n}(y)} {mg}\ \end{equation}% -----------------------------------------------------------------------------------------
are the pressure scale heights of ions ($\Lambda_{i}$) and neutrals ($\Lambda_{n}$), and $p_{0 \, i,n}$ denotes the gas pressure of ions and neutrals at the reference level, $y=y_{\rm r}=10$ Mm.  In  Equation~(\ref{eq:lambda}), $T_{i}(y)$ and $T_{n}(y)$ stand for temperatures of ions and neutrals, respectively. These temperatures are taken from the model of \cite{Avrett2008} 
%and they are illustrated in 
(see Figure~\ref{fig:1}). Note that for $y>0.2$ Mm the temperature of ions is higher than the temperature of the neutrals. In the photosphere, which is localized at $0 \leq y\leq 0.5$ Mm, $T_{i}(y)$ increases with $y$ until close to the bottom of the chromosphere, specifically for $y \approx 0.7$~Mm, $T_{i}(y)$ reaches its local maximum of about $1.5 \cdot 10^{5}$ K. Higher up $T_{i}(y)$ decreases with height until at the TR it rises abruptly to the coronal value of about 1~MK. The neutral temperature attains its value of about 5 kK within the region of $0 \leq y \leq 1.75$ Mm. It decreases suddenly right below the TR attaining its minimum of about $150$~K and higher up it increases with $y$ reaching its coronal magnitude of about 0.8 MK at $y = 10$ Mm. 
%The average plasma temperature that results from ion and neutral temperature profiles, is given as    
% -----------------------------------------------------------------------------------------
%\beq
%\label{eq:avtemp}
%T(y)=\frac{\varrho_{n}(y)}{\varrho_{n}(y)+\varrho_{i}(y)}T_{n}(y)+\frac{\varrho_{i}(y)}{\varrho_{n}(y)+\varrho_{i}(y)}T_{i}(y)  \, .
%\eeq
% -----------------------------------------------------------------------------------------
%and is displayed in Fig.~\ref{fig:2}. 

As a result of the adopted temperature profiles, the mass densities of ions and neutrals decline with height. The left panel of Figure~\ref{fig:3} shows the ratio of the mass density of ions to the mass density of neutrals, $\varrho_{i}(x,y)/\varrho_{n}(x,y)$, which result from Eq.~(\ref{eq:pres}). 
%and ion and neutral profiles displayed in Fig.~\ref{fig:1}. 
Note that below the level of $y=2.1$ Mm the plasma is dominated by neutrals with a minimum of $\varrho_{i}/\varrho_{n}\approx10^{-4}$, while the corona consists of essentially fully ionized plasma; we added a small amount of neutrals in the solar corona due to numerical reasons. We specify the plasma $\beta$ as the ratio of ion plus neutral gas pressures to magnetic pressure,
% -----------------------------------------------------------------------------------------
\beq
\label{eq:plasmabeta}
\beta(y)=\frac{p_{i}(y)+p_{n}(y)}{B^{2}(y)/2\mu} \, .
\eeq
% -----------------------------------------------------------------------------------------
The spatial profile of plasma $\beta$ is illustrated in the right panel of Figure~\ref{fig:3}. Note that within the displayed region for the coronal plasma, for $y>2.1$ Mm, and along the entire $x=0$ line $\beta$ is smaller than 1.

% -----------------------------------------------------------------------------------------
\subsubsection{Perturbations}\label{sec:pert}
% -----------------------------------------------------------------------------------------

Initially, at $t=0$~s, we perturb the model equilibrium by launching, at the bottom boundary, simultaneously time-dependent signals in the 
%$y-$components of ion and neutral velocities, and 
ion and neutral gas pressures, which are expressed as follows:

\beqa
\begin{split}
%\label{eq:velocity}
%\left[V_{iy},V_{ny}\right](x,y,t=0) = A_{\rm V}\left[1,1\right] \times {\rm exp}\left(-\frac{x^{2}+(y-y_{0})^{2}}{w^{2}}\right) \, , \\  \left[i,n\right]
\label{eq:pressure}
\begin{aligned}
&p_{\left[{\rm i}, \, {\rm n}\right]}(x,y,t=0) \\
&=p_{0}\left( 1+A_{p \, [{\rm i,n}]} \, {\rm exp}\left(-\frac{x^{2}+(y-y_{0})^{2}}{w^{2}}\right)\, f(t)\right) \, , \\
&f(t)=\begin{cases}
               \,  1-{\rm exp}(-t/\tau)  \, , \, \, \, \, \, \, \, \, \, \, \, \, \, \, \, \, \, \, \, t\leq \tau_{\rm max} \\
			   \,  {\rm exp}(-(t-\tau_{\rm max})/\tau)\, ,  \,  \,  \,  \,  \,  \,  \,  \,  t> \tau_{\rm max}
            \end{cases} \, . \\
            \end{aligned}
\end{split}
\eeqa
Here $y_{0}$ is the vertical position of the signals, $w$ is their width 
%and $A_{\rm V}$ 
and $A_{\rm p\, i}$, $A_{\rm p\, n}$ their amplitudes. Function $f(t)$ denotes the temporal profile of the gas pressure signal, with $\tau$ being its characteristic growth/decay time and $\tau_{\rm max}$ time at which $f(t)$ reaches its maximum. We set and hold fixed $w=0.1$ Mm, $\tau=50$ s and $\tau_{\rm max}=30$ s, while allowing other parameters to vary. For our studies, the signal position, $y_{0}$, varies between 1.65 Mm and 1.85 Mm 
%the amplitude $A_{\rm V}$ changes between 30 km s$^{-1}$ and 50 km s$^{-1}$ and 
and the amplitude $A_{\rm p}$ between 6 and 10. This signal position corresponds to a region of $\beta<1$ (Figure~\ref{fig:3}, right). As a result of that, slow and fast magnetoacoustic waves are weakly coupled there and the initial pulse triggers slow waves, which propagate essentially along magnetic field lines (e.g. \citealt{Nakariakov2005}). Unless otherwise stated, the further analysis focuses on the case for $A_{\rm p}=8$ 
%$A_{\rm V}=35$ km s$^{-1}$ 
and $y_{0}=1.75$ Mm, which corresponds to plasma heated to maximum temperature of 60 kK 
%that is moving upwardly with velocity of 35 km s$^{-1}$ 
$1.75$ Mm above the photosphere. %(SOME CORRESPONDENCE WITH IRIS OBSERVATIONS MAY BE DONE HERE)} 
We have verified that only such strong pulses launched at the top of the chromosphere result in spicules. Lower amplitude pulses excite smaller jets and pulses launched from the lower chromospheric layers lead to horizontally spread jets. The time-dependent signal launched at $y_{0}=1.75$ Mm may mimic a post-reconnection event.

% -----------------------------------------------------------------------------------------
\section{Numerical Simulations of Two-fluid Equations}\label{sec:num_sim_MHD}
% -----------------------------------------------------------------------------------------
To solve two-fluid equations numerically, we use the JOANNA code (\citealt{Woj2017}, in preparation). In our problem, we set the Courant--Friedrichs--Lewy number (\citealt{Courant1928}) equal to 0.3 
%%%%%%%%%%%%%%%%%%%%%%%%%%%%%%%%%%%%%%%%%%%%%%%%%%%%%%%%%%%%%%%%%
\begin{figure*}[!ht]
\centering
\mbox{
%\hspace*{-1.5cm}
\includegraphics[scale=0.4, angle=0]{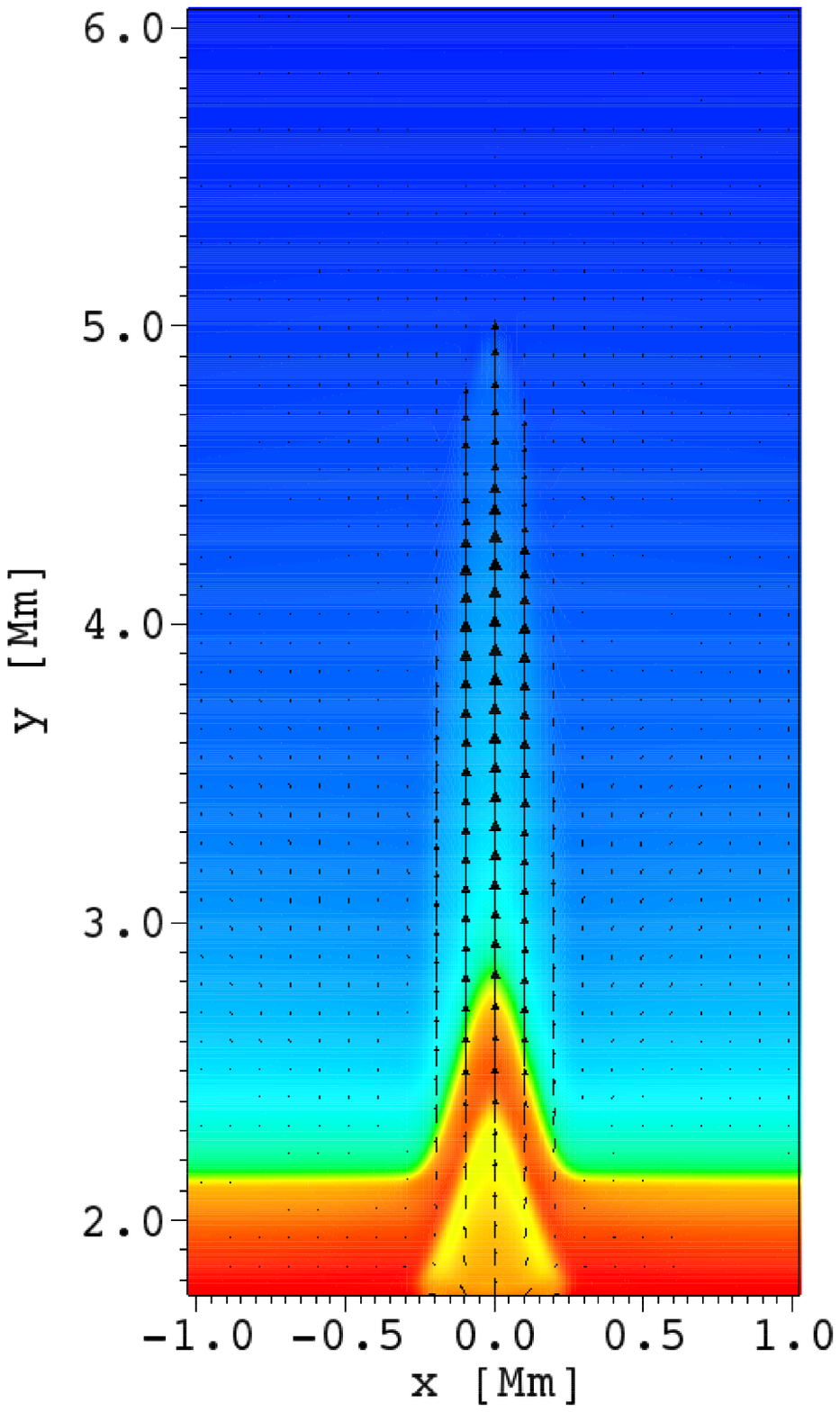}\hspace*{-3.5cm}
\includegraphics[scale=0.4, angle=0]{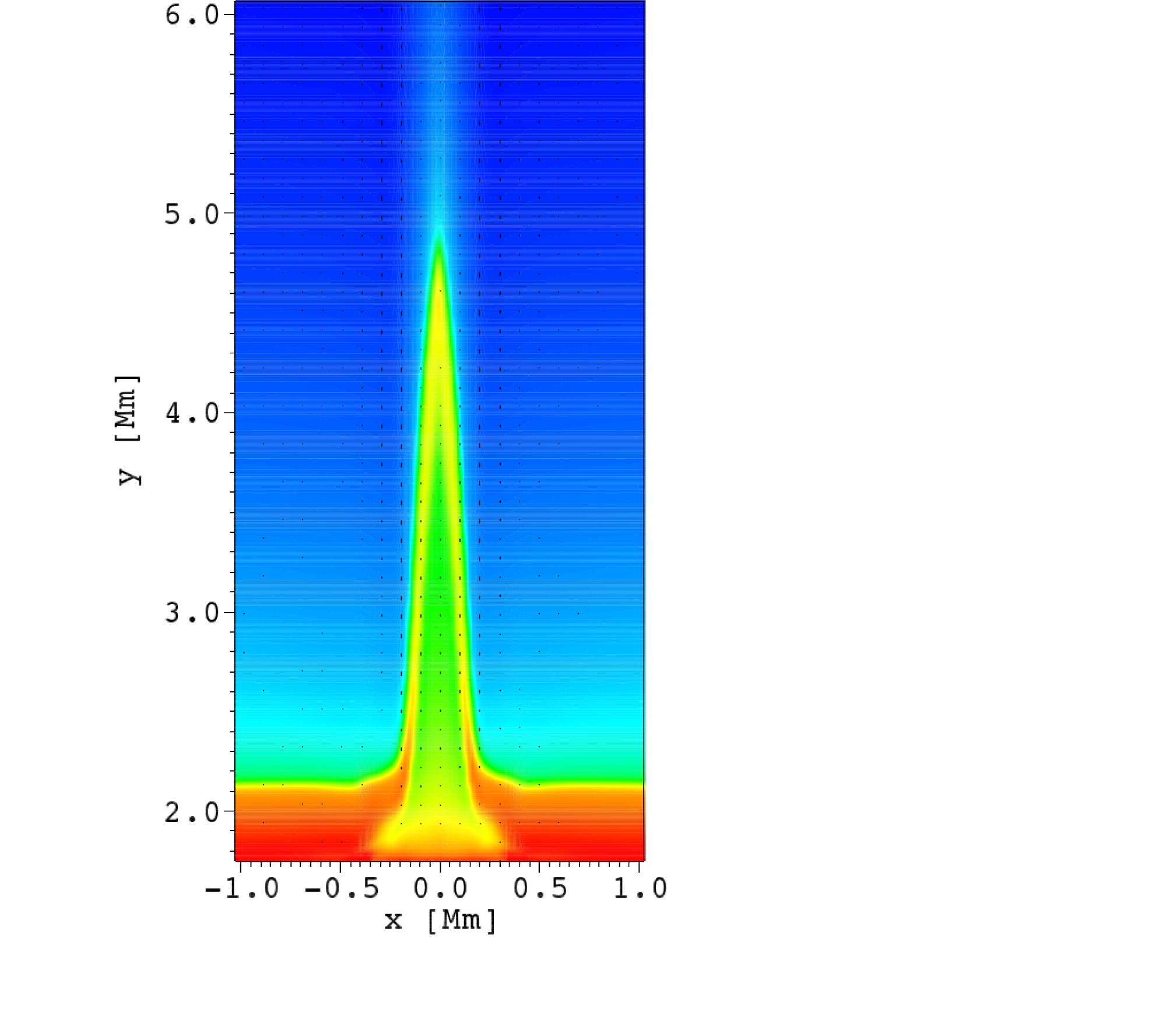}\hspace*{-3.5cm}
\includegraphics[scale=0.4, angle=0]{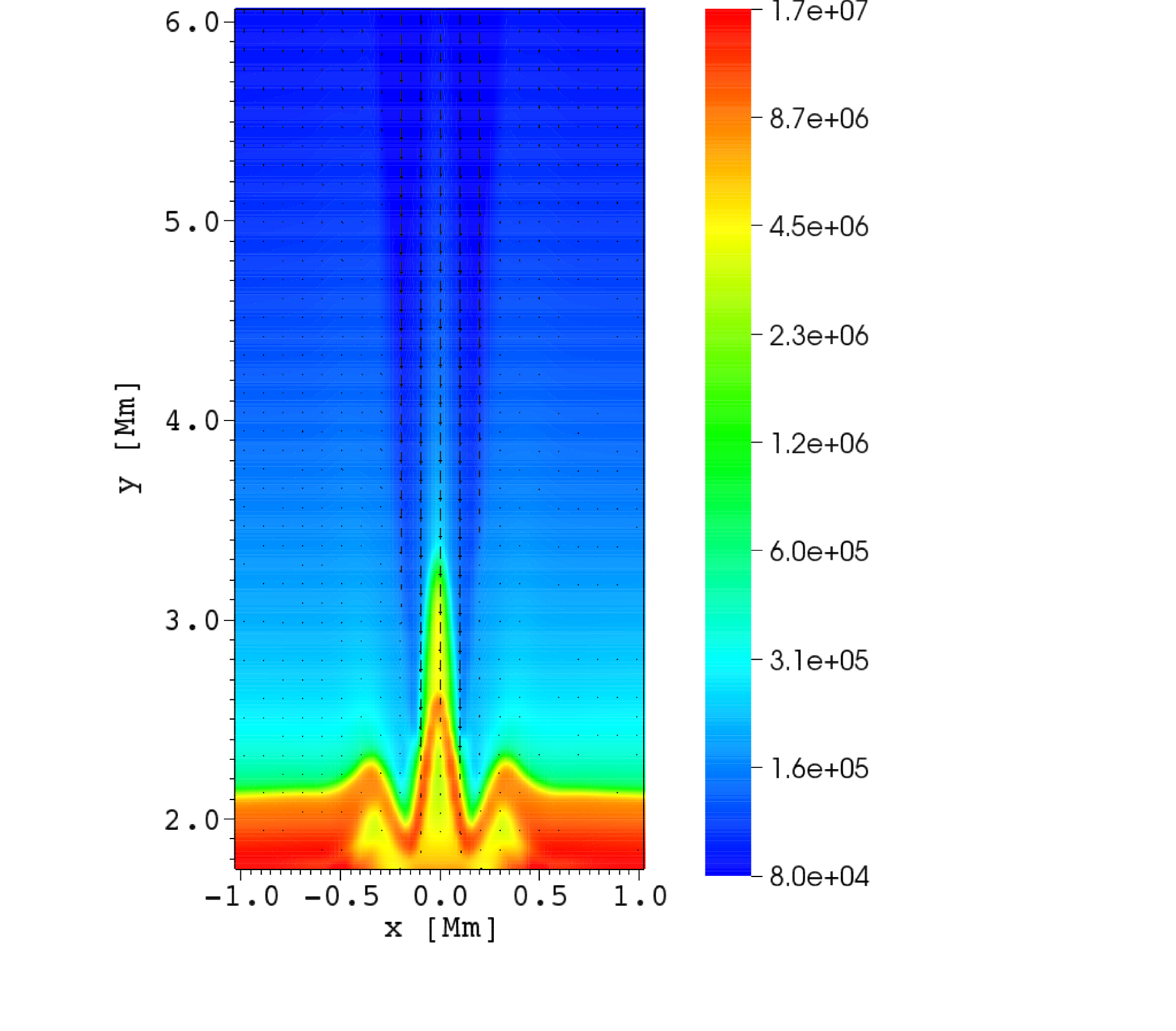}
}
\mbox{
\vspace*{-2.5cm}
%\hspace*{-1.5cm}
\includegraphics[scale=0.4, angle=0]{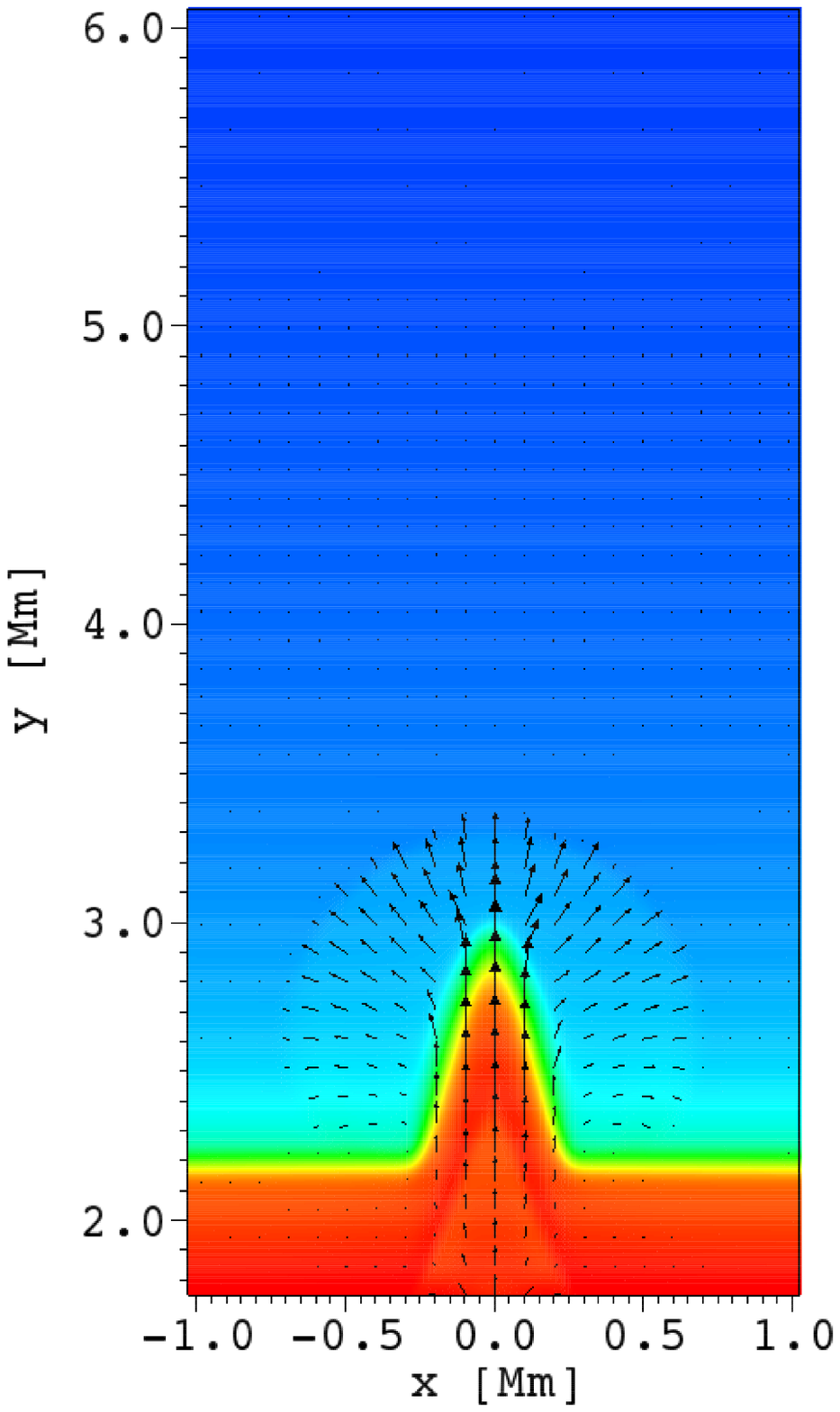}\hspace*{-3.5cm}
\includegraphics[scale=0.4, angle=0]{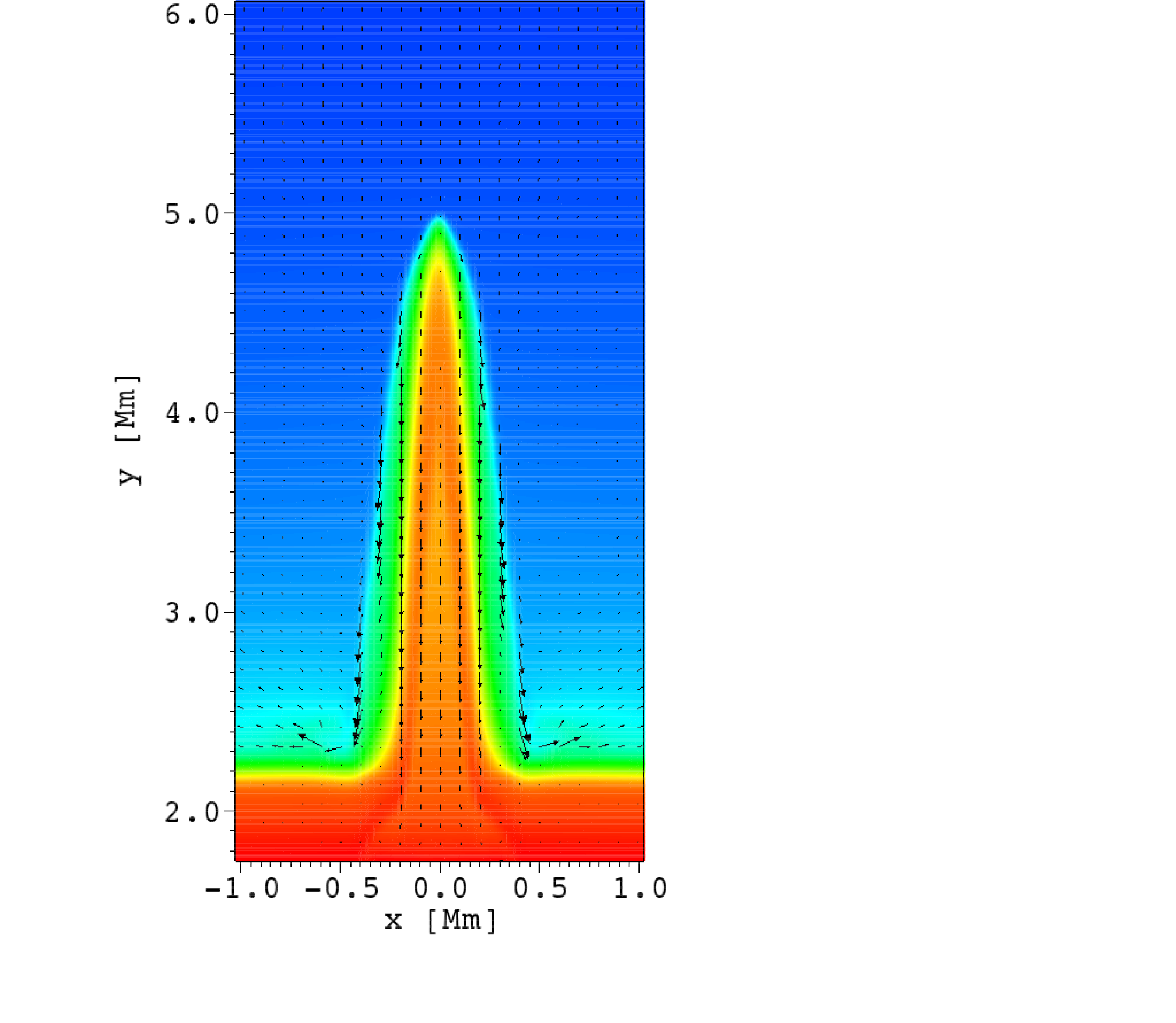}\hspace*{-3.5cm}
\includegraphics[scale=0.4, angle=0]{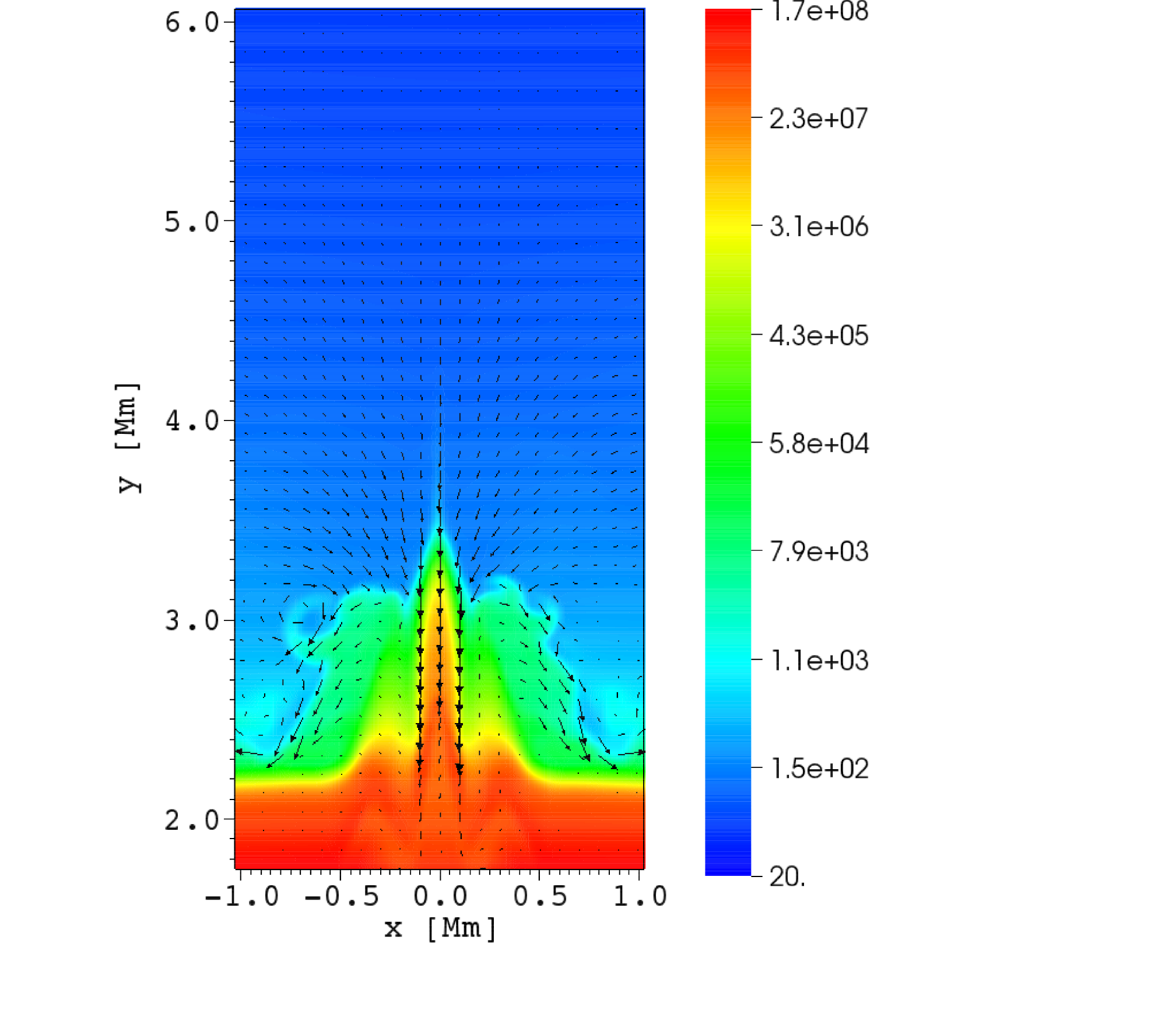}
}
%\vspace{1.4cm}
\caption{Temporal evolution of ${\rm log} (\varrho_{i}(x,y))$  
at $t=70$ s, $t=210$ s, and  $t=320$ s (from top left to top right),
and ${\rm log} (\varrho_{n}(x,y))$ (from bottom left to bottom right). 
Arrows represent ion and neutral velocity vectors in the $x-y$ plane. Hot coronal plasma is blue, cold chromospheric plasma is orange-red, and intermediate transition region plasma/interface between them is green on the colormap.
}
\label{fig:4}
\end{figure*}
%%%%%%%%%%%%%%%%%%%%%%%%%%%%%%%%%%%%%%%%%%%%%%%%%%%%%%%%%%%%%%%%%%%%%%%%%%%%%%%%%%%%%%%%%%%%%%%%
and specify the simulation box in $(x,y)$ as $(-1.28, 1.28)$~Mm $\times$ $(1.75, 50.0)$ Mm, where $y=0$ denotes the bottom of the photosphere. In the numerical simulations we adopt the uniform grid within the region $(-1.28 \leq x \leq 1.28)$ Mm $\times$ $(1.75 \leq y \leq 6.87)$ Mm, which is covered by $256 \times 512$ grid points. This grid leads to a resolution of $10$ km in the main region of the simulation box that is below $y=6.12$ Mm. Above this region, namely within the rectangle $(-1.28 \leq x \leq 1.28)$ Mm $\times$ $(6.87 \leq y \leq 50.0)$ Mm, we implement a stretched grid along the $y-$direction; this box is divided into 128 cells whose size grows with $y$. Such a stretched grid plays the role of a sponge as it absorbs incoming signal and minimizes reflections from the top boundary. We impose open boundary conditions for outflowing signal at the side boundaries, while at the top and bottom we fixed all plasma quantities to their equilibrium values.

We launch the gas pressure 
%and velocity 
signals in the chromosphere varying their initial position, $y_{0}$, and amplitudes, 
%$A_{\rm V}$, 
$A_{\rm p \, i}$ and $A_{\rm p \, n}$, as described in Section~\ref{sec:pert}. The basic mechanism behind the simulations is that the signals, while launched from the region of $\beta < 1$, essentially split into two, counter-propagating along the equilibrium magnetic field lines' slow magnetoacoustic (slow henceforth) waves; downwardly propagating slow waves decay in time (not shown in the framework of these simulations), while upwardly moving slow waves grow in their amplitudes. The latter convert into a slow shock at higher altitudes due to the decrease of mass density with height. The chromospheric plasma lags behind the shock front to form a contact wave consisting of a spicule. The simple waves structure associated with this phenomenon is described by \cite{Kuzma2017}. Note that the signal in $A_{p \, i}$ excites fast magnetoacoustic waves too but they are of low amplitudes as they spread quasi-isotropically in space. The pulse in $A_{p \, n}$ generates neutral acoustic waves (\citealt{Zaqarashvili2011}).

The top row of Figure~\ref{fig:4} shows the spatial profiles of log($\varrho_{i}(x,y)$) at three instants of time, which are $t=70$ s (top left), $t=210$ s (top middle) and $t=320$ s (top-right). The gas pressure drivers of Eq.~(\ref{eq:pressure}) operate at the point of ($x=0,y=y_{0}=1.75$)~Mm, which is located in the chromosphere around $0.35$~Mm below the TR, and they reach their maximum at $t=30$ s. 
%The amplitudes of the initial pulses \textbf{is $A_{\rm p}=8$.} 
At $t=70$ s, it is found that the ion shock front is followed by hot plasma, which results from the drivers, and it reaches a level of $y\approx5.0$~Mm. It is followed by a second, stronger shock, which results at maximum of $f(t)$. However, the chromospheric (cold) ion plasma is located at $y\approx2.8$~Mm at this time (see Figure~\ref{fig:4}, the top-left panel). The shock fronts move continuously upward, which attain the level of more than $y=25$~Mm (not drawn here) and the chromospheric plasma reaches its maximum height of $y\approx4.9$~Mm at $t= 210$ s (Figure~\ref{fig:4}, the top-middle panel). 
%The perturbations propagate in both directions (i.e., up and down) from the pressure launching site (cf., Fig.~\ref{fig:4}, top/bottom left-panels). However, the downward propagating perturbations die soon due to the more denser region towards the photosphere. 
The head of the spicule remains denser (yellow color on the colormap), while its interior rarefies in time (green color on the colormap) due to rarefaction wave propagating upwards. Above the apex of the ion spicule, we can see a constant stream of ions injected into higher layers of the solar corona. 
In addition, the down-falling of the ion spicule plasma also starts as the time progresses. The reverse velocity arrows in the central region and sides of the ion spicule justifies downfall of the cold chromospheric plasma. 
%, which creates the V-shaped structure near the base (top right-panel). 
This downfall is more stronger during the decay phase (see reverse velocity arrows in the top-middle and top-right panels), which creates the V-shaped structure as the down-falling velocity is non-uniform along the horizontal direction (see the top-right panel). 
%The V-shaped structure is not visible during the maximum phase of neutral spicule (bottom middle-panel), however, 

%\textbf{In the decay phase, the downfalling of the spicule plasma (see; reverse velocity arrows) is clearly visible, which supress the the ion spicule upto the height of $y=3.2$~Mm (cf., Fig.~\ref{fig:4}, the top right-panel). In the similar fasion, neutral spicule also suppressed upto the height of $y=3.7$~Mm (cf., Fig.~\ref{fig:4}, the bottom right-panel)  
%In the meantime,   Therefore, this V-shaped structure forms due to the downfalling of the plasma near the base of the ion spicule. 

%Below the spicule apex, at the height of $y=3.8$ Mm, we spot the moving upward rarefaction wave. Additionally we notice, that the plasma at the base of the spicule begins to fall toward the lower layers of the solar atmosphere. The top right-panel, which corresponds to $t=140$~s, shows that the chromospheric plasma is located at its maximum height of $y\approx 5.4$~Mm, and the shock front has gone well beyond the displayed region.  

%%%%%%%%%%%%%%%%%%%%%%%%%%%%%%%%%%%%%%%%%%%%%%%%%%%%%%%%%%%%%%%%%%%%%%%%%%%%%%%%%%%%%%%%%%%%%%%%%
\begin{figure*}[!ht]
\centering
\mbox{
\hspace*{-1.2cm}
\includegraphics[scale=0.4, angle=0]{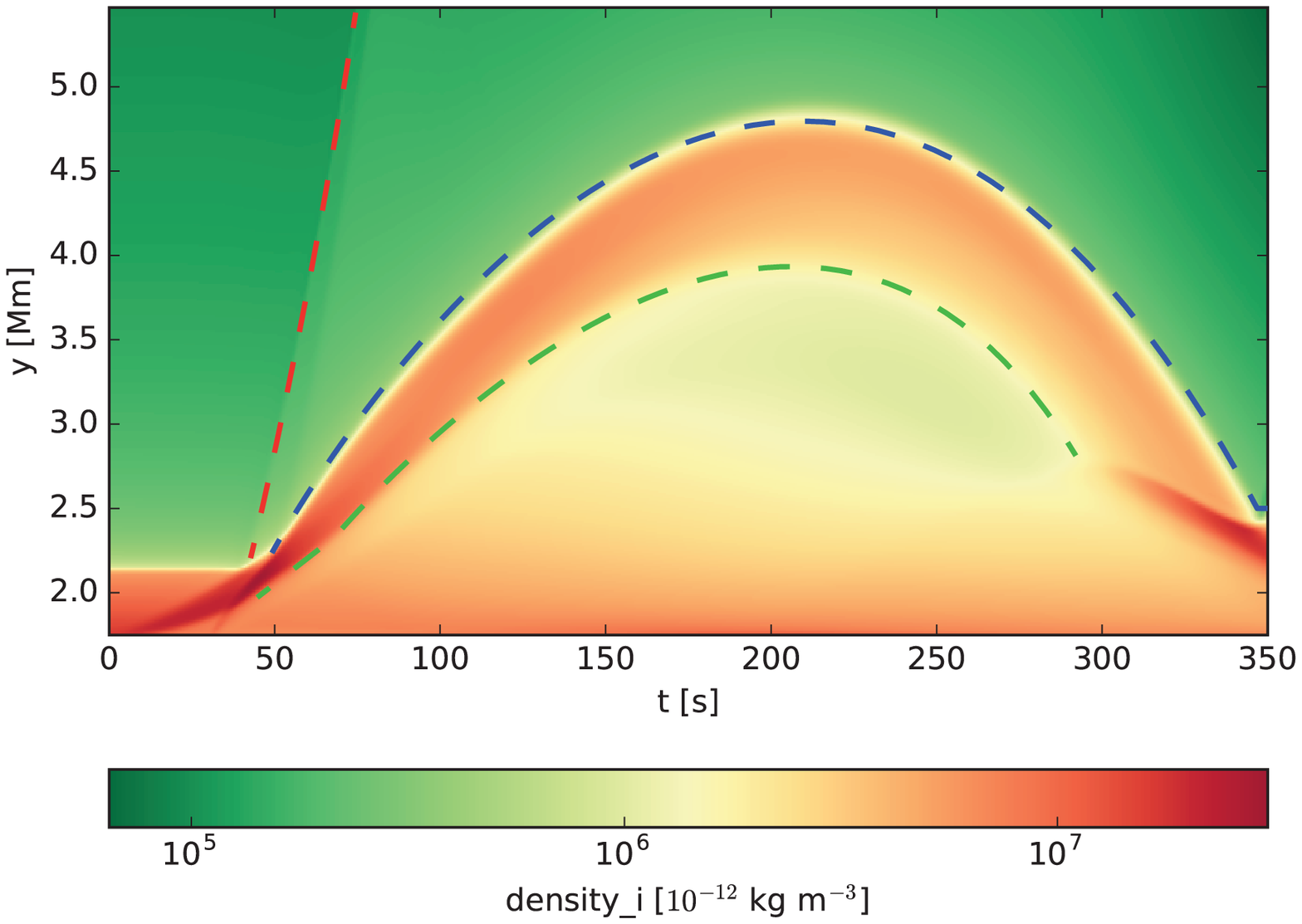} \hspace*{-0.8cm}
%\vspace{-0.75cm}
\includegraphics[scale=0.4, angle=0]{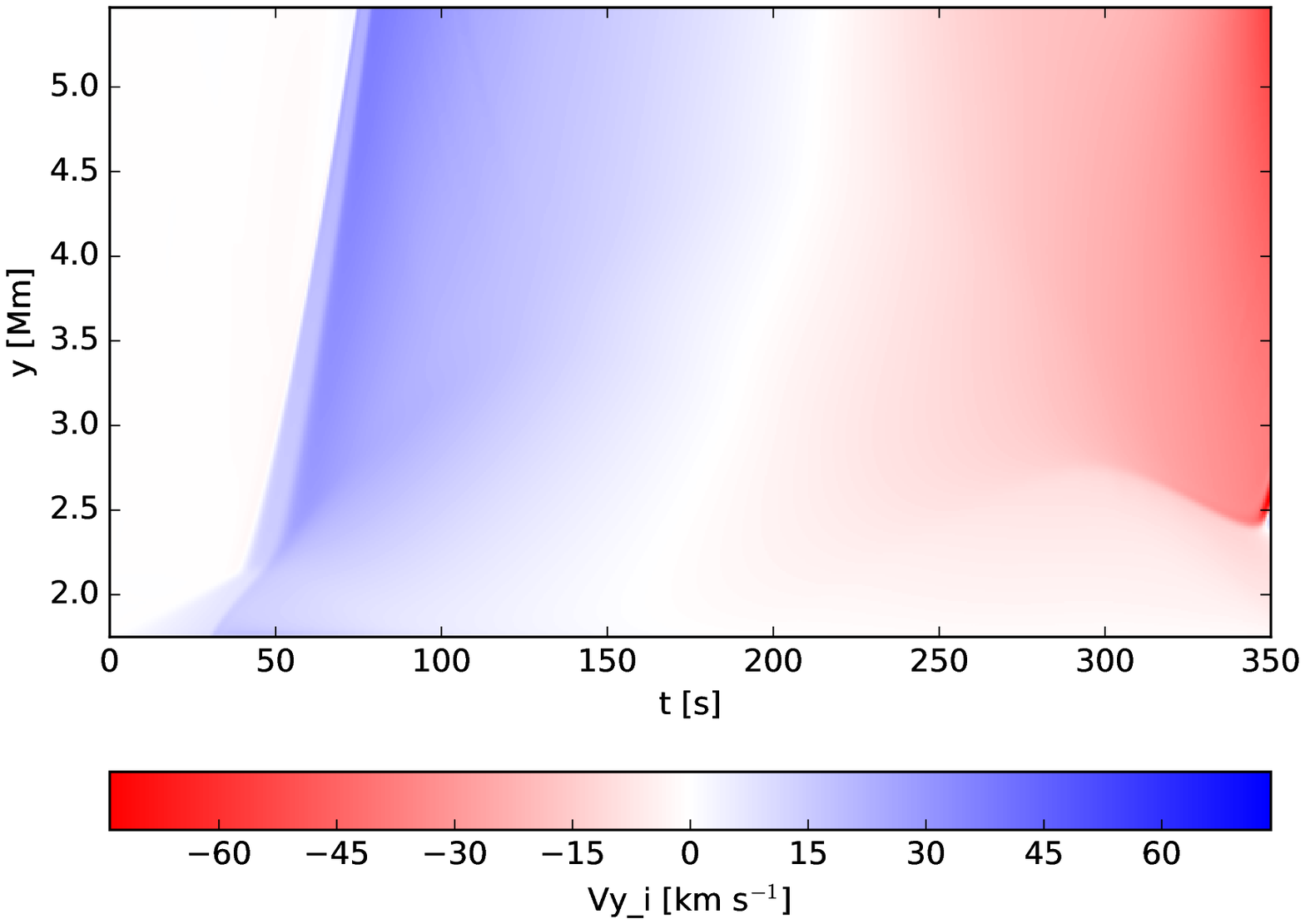}
}
\mbox{
\hspace*{-1.2cm}
\includegraphics[scale=0.4, angle=0]{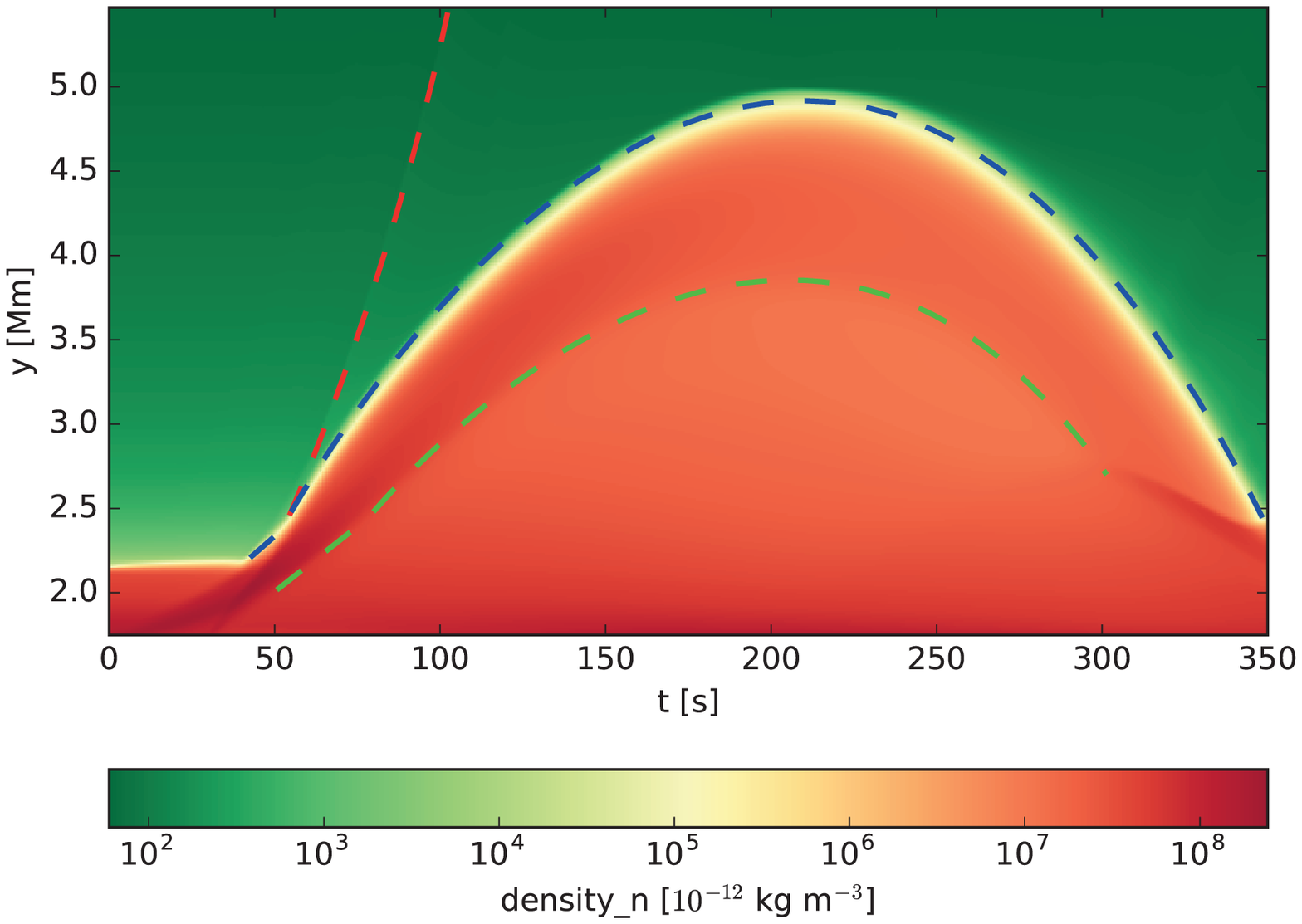}\hspace*{-0.8cm}
\includegraphics[scale=0.4, angle=0]{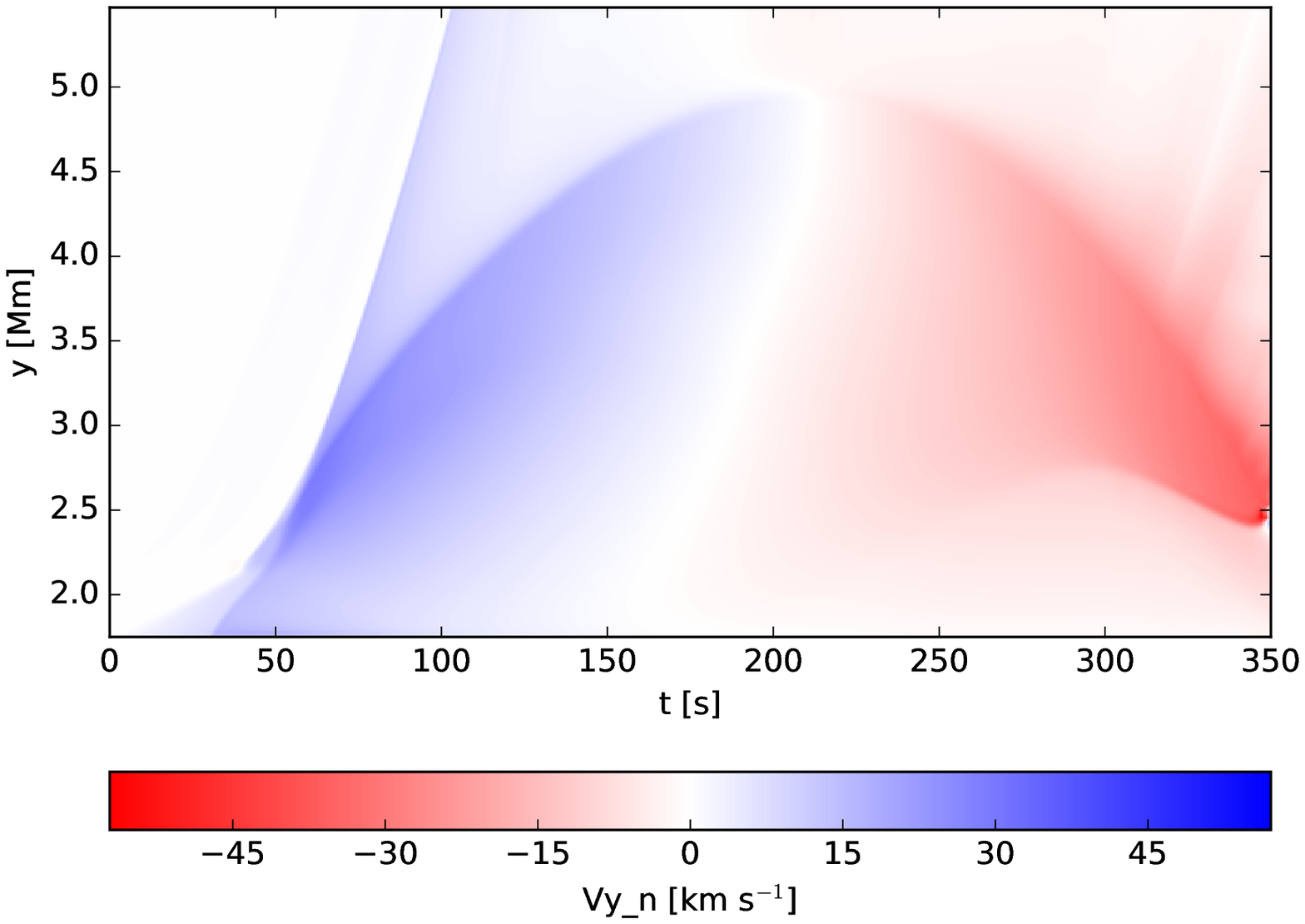}
%\vspace{-1.75cm}
}
%\vspace{1.4cm}
\caption{Temporal evolution of log($\varrho_{i,n}(x=0,y)$) (left) and $V_{i,n \, {\rm y}}(x=0,y)$ (right) 
%in the case of $A_{\rm V}=40$ km s$^{-1}$ and $y_{0}=1.75$ Mm
for the ions (top) and neutrals (bottom). The red, blue, and green dashed lines indicate slow shock, contact wave, and rarefaction wave, respectively. Transition from the red to the green area shows the contact waves.
}
\label{fig:6}
\end{figure*}
%%%%%%%%%%%%%%%%%%%%%%%%%%%%%%%%%%%%%%%%%%%%%%%%%%%%%%%%%%%%%%%%%%%%%%%%%%%%%%%%%%%%%%%%%%%%%%

We have implemented two-fluid approach in our numerical experiment to understand the behavior of ions and neutrals in the spicule dynamics. It may be possible that the ions and neutrals may exhibit the different dynamics. Therefore, we show the corresponding profiles of log ($\varrho_{n}(x,y)$) in the bottom panels of Figure~\ref{fig:4}. The overall dynamics of the neutrals (i.e., formation of a shock front and lagging off chromospheric plasma) are qualitatively similar to the dynamics of ions in this numerical experiment. However, the slow neutral shock does not propagate along magnetic field lines, forming a circle-like front (the left bottom panel). There is also no upward stream consisting of neutrals above the apex of the ion spicule (the bottom-middle panel). The spicule is surrounded by the neutral cloud in which we can spot vortices of Rayleigh-Taylor instabilities during decay phase. The down-falling of neutral gas is very weak during maximum phase (the middle right-panel), which is mostly dominated towards the edges of the neutral spicule. However, strong down-flow of neutral gas occurs during the decay phase. Neutral spicule has V-shaped structure resemblance in the decay phase (bottom-right panel) but not as prominent as it occurs in the ion spicule. As neutrals are not guided by the magnetic field, the neutral gas experiences more horizontal spread compared to the ions. So, the neutral spicule does not exhibit very sharp V-shaped structure due to the dominance of horizontal spreading of neutrals.   

In addition, the top of the ion (neutral) spicule is suppressed till the height of $y=3.4$~Mm ($y=3.5$~Mm) due to the dominance of down-falling in the decay phase. This down-falling is strongest on sides of the ion/neutral spicule and decrease outwards, which creates the V-shape structure at the bottom of ion/neutral spicules. Most of the chromospheric plasma, which was injected into the corona falls toward the chromosphere/TR. 

We estimate now the width of the spicule in neutrals and ions at $y=3.0$~Mm during the maximum phase of the spicule. By looking at horizontal mass density profiles and assuming that the mass density above (below) 1.2 times of the background mass density (i.e., mass density in the absence of any spicule; at $x=-1.0$ Mm) are considered as the starting (end) points of the spicules. We find that the neutral spicule is wider $(\sim 600$ km) than the spicule consisting of ions $(\sim 400$ km).

\begin{figure*}[!ht]
\centering
\mbox{
\hspace*{-0.7cm}
\includegraphics[scale=0.25,  angle=0]{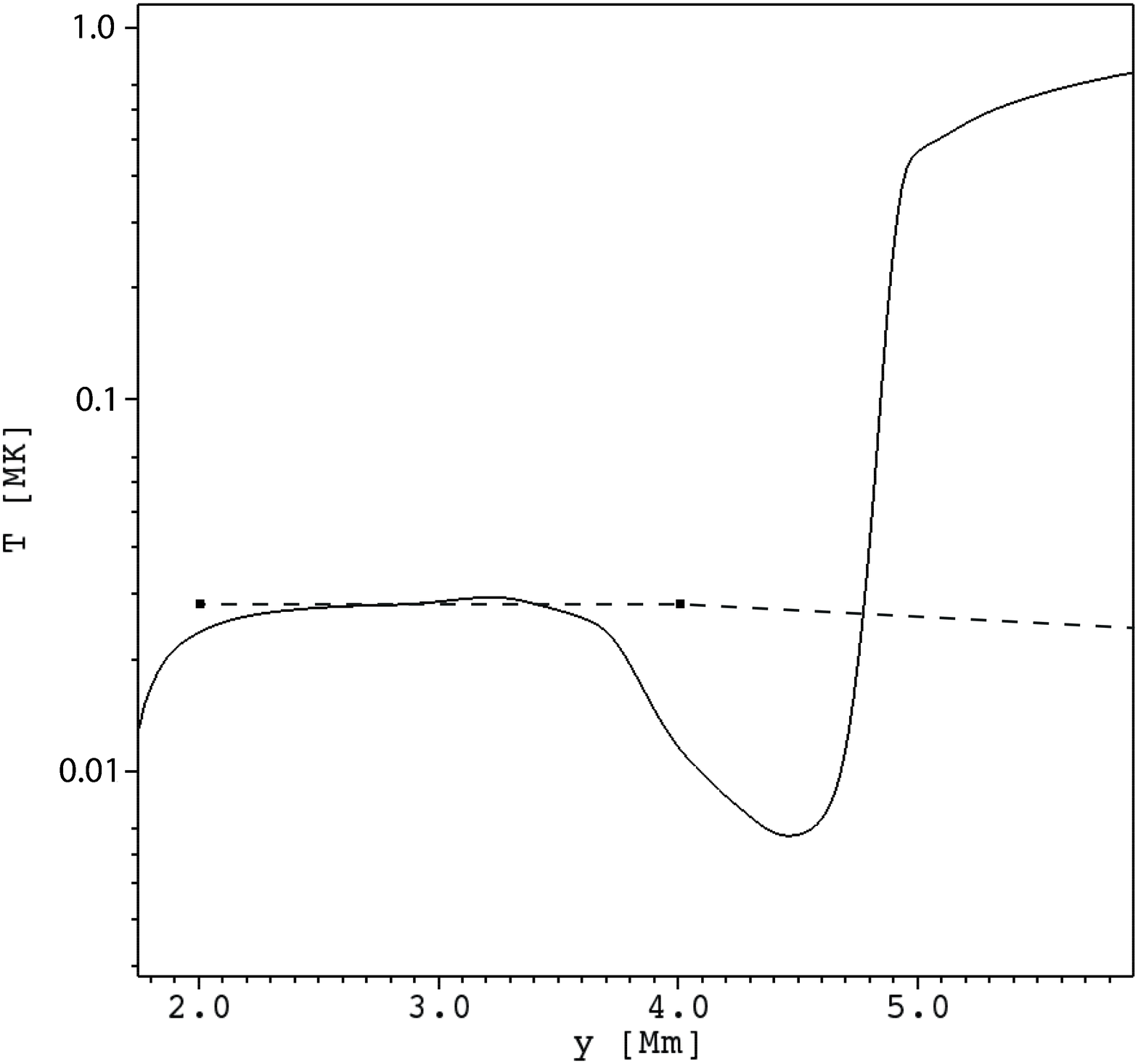} %\hspace*{-0.8cm}
%\hspace{1.0cm}
\includegraphics[scale=0.25,  angle=0]{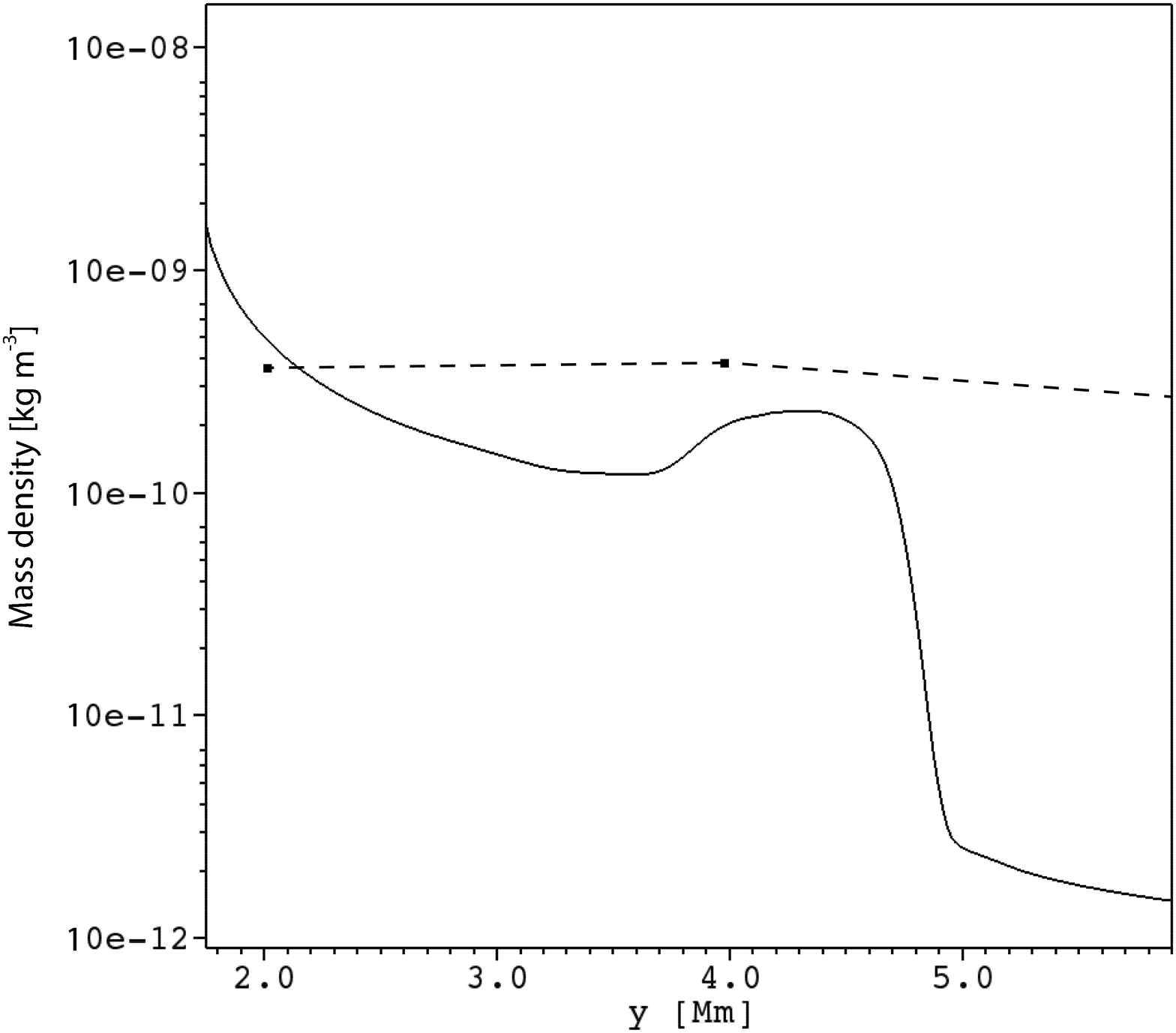}
}
%        \begin{center}
%        \mbox{\hspace{-2.0cm}
%                \includegraphics[scale=0.25, angle=0]{./figs/temperature.eps}
%                }
%\vspace{-0.5cm}
                \caption{Vertical profiles of average temperature (the left panel) and mass density (the right panel) of the spicule in the simulation evaluated along $x=0$~Mm, at $t=210$~s (the solid lines) and observational data of \cite{Beckers1968} (the dashed lines). 
                %for $A_{\rm V}=35$ km s$^{-1}$, $A_{\rm p}=2$ and $y_{0}=1.75$ Mm 
%. The dashed line represents average temperature of ambient plasma.  
}
                \label{fig:51}
%        \end{center}
\end{figure*}
%%%%%%%%%%%%%%%%%%%%%%%%%%%%%%%%%%%%%%%%%%%%%%%%%%%%%%%%%%%%%%%%%%%%%%%%%%%%%%%%%%%%%%%%%%%%%%%%%%%

We discuss now temporal evolution of mass densities and vertical velocities of ions and neutrals. 
%for a case of $A_{\rm V} = 40$ km s $^{-1}$ and $y_{0}=1.75$ Mm. 
Figure~\ref{fig:6} displays time signatures of $\varrho_{i,n}(x=0,y,t)$ (the left panels) and $V_{i,n \,{\rm y}}(x=0,y,t)$ (the right panels) for ions (the top panels) and neutrals (the bottom panels). The rise time of the chromospheric ions to their maximum height is about $210$~s (see Figure~\ref{fig:6}, top-left). In the temporal evolution profile of the ions, we spot three major waves above the TR (the top-left panel): the leading wave is a slow shock wave (red line) that is followed by the contact wave (blue line) and the rarefaction wave (green line). The contact wave does not appear in the top-right panel since there is no jump in velocity of ions across a contact surface with normal almost perpendicular to ${\bf B}$. See also \cite{Kuzma2017} for a similar discussion in the MHD case. The light-green area between slow shock and contact wave reveals that significant amount of ions are constantly injected above the spicule apex during its lifetime. This phenomenon is absent in the case of neutrals (the bottom-left panel). The ion slow shock travels with higher speed ($\sim$150 km s$^{-1}$) than neutral slow shock ($\sim$90~km~s$^{-1}$; see the right panels). Both ions and neutrals injected into the corona above the spicule apex accelerate with height. 
The light-green area below the apex of the ion spicule reveals that the rarefaction wave exerts a greater impact on the already rarefied ion spicule. The boundary between chromospheric and coronal neutrals is also sharper than between chromospheric and coronal ions. This is a result of merging between the top of the ion spicule and highly ionized corona. From the temporal evolution profile of the neutrals, we infer that both shock and rarefaction waves are of low amplitudes, to the point, when the rarefaction wave is almost unnoticeable (green line, the bottom-left panel). At a later time, the chromospheric plasma, which was earlier injected into the corona, begins to fall toward the TR.

%%%%%%%%%%%%%%%%%%%%%%%%%%%%%%%%%%%%%%%%%%%%%%%%%%%%%%%%%%%%%%%%%%%%%%%%%%%%%%%%%%%
\begin{figure*}[!ht]
\centering
\mbox{
\hspace*{-0.7cm}
\includegraphics[scale=0.45,  angle=0]{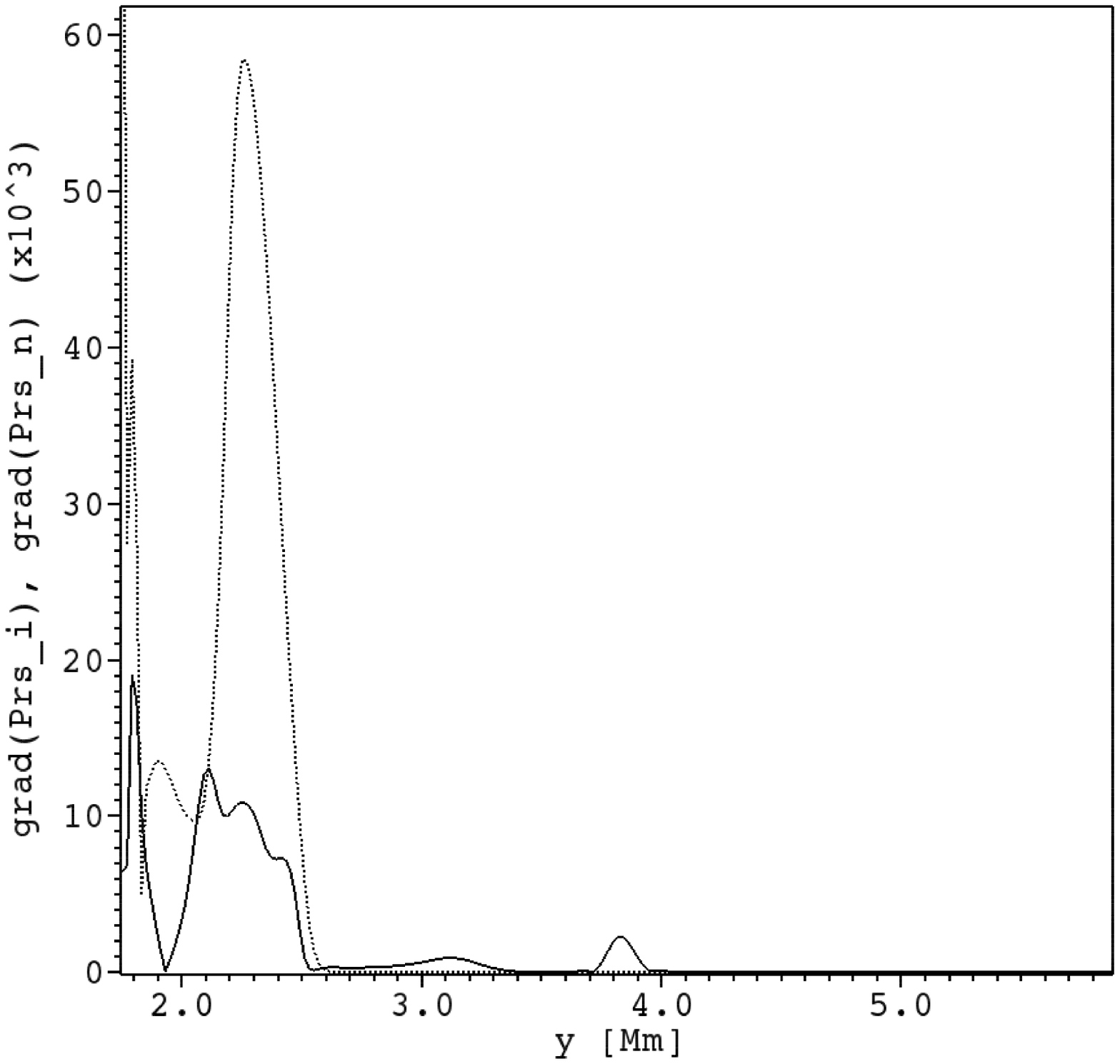} %\hspace*{-0.8cm}
%\hspace{1.0cm}
\includegraphics[scale=0.45,  angle=0]{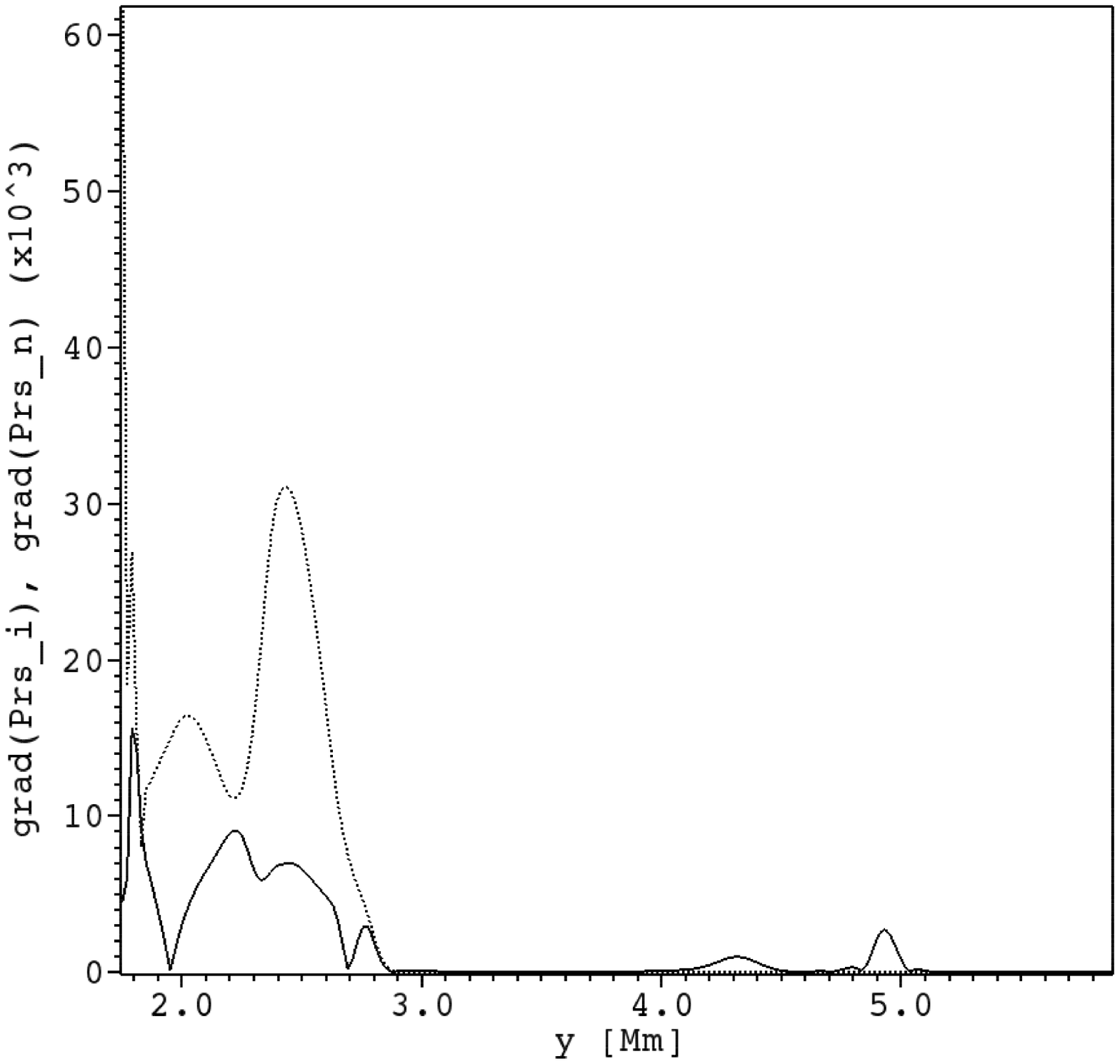}
}
\caption{Vertical profiles of the gas pressure gradient of neutrals (the dotted line) and ions (the solid line) 
%for $A_{\rm V}=35$ km s$^{-1}$, $A_{\rm p}=2$ and $y_{0}=1.75$ Mm 
evaluated along $x=0$~Mm, at $t=60$~s (the left panel) and $t=70$~s (the right panel).  
}
\label{fig:50}
\end{figure*}

The spicule consists mostly of dense and cold neutrals. The ionization level at all stages of the spicule evolution remains at about $10\%$. Figure~\ref{fig:51} (left) shows the vertical profile of the average temperature of the simulated spicule (solid line) along with the widely used observed values (dashed line; \citealt{Beckers1968}). The average temperature of the spicule (solid line) is 100 times lower than the coronal plasma, which shows a constant value of temperature up to a height of $\sim$ 2.8 Mm. The spicule plasma exhibits its lowest temperature of 9000 K in the upper part of the spicule, $\sim 4 500$ km above the TR. Higher up, the temperature experiences an abrupt jump at the contact wave, reaching temperatures of up to 800,000 K and matching the local coronal temperature. Figure~\ref{fig:51} (left) shows the vertical profile of mass density of the spicule (solid line). The cold head of the spicule is about twice denser than its core, which is resembled by the weak bump around $y=4.2$ Mm. Above the apex of the spicule, mass density decreases to its coronal values. The temperature as well as mass density of the simulated spicule are compared with the classical results of the spicule (dashed line in both-panels of Figure~\ref{fig:51}; \citealt{Beckers1968}). Observationally, it is reported that mass density and temperature are almost constant over the whole length of the spicule, which vary by about 10 \% over the whole length of the spicule (\citealt{Beckers1968}). However, the simulated spicule exhibits a significant variations in the temperature and mass density along its length. It should be noted that the temperature of the spicule matches with the observed values over a height range of 2.0~Mm~$<y<$~3.7~Mm.
 %which stays correct compared to observations (\citealt{Beckers1968}, dashed line).      

We now discuss gradients of gas pressures of neutrals and ions. The corresponding vertical profiles are displayed in Figure~\ref{fig:50}. The dotted line shows the gas pressure gradient of neutrals, while the solid line illustrates the gas pressure gradient of ions along the $y$-direction. The left (right) panel illustrates the vertical distribution of the gas pressure gradient at $t=60$ s ($t=70$ s). It is clear that the neutral gas pressure gradient (the dotted line) is very high in comparison to the ion gas pressure gradient (the solid line) at these instants of time. The difference between neutrals and ions pressure gradients is higher at about $y=1.75$ Mm at which the driver in gas pressures operates. The gradients approach each other far away from the driver. The significant gas pressure gradient difference between the ions and neutrals is an important result, which clearly leads to a different dynamics of ions and neutrals. % {\bf and ...}.    
% -----------------------------------------------------------------------------------------
\section{Discussions and Conclusions}\label{sec:Summary}
% -----------------------------------------------------------------------------------------
We have performed numerical simulations of a spicule by setting in the upper chromosphere localized time-dependent signals in ion and neutral gas pressures. 
%and $y$-components of ion and neutral velocities 
The initial magnetic field configuration was current-free, and the atmosphere was stratified hydrostatically. The whole physical system was described by a set of two-fluid equations. Our numerical findings revealed that, as a result of the rapid decrease of the mass density with height, an upwardly propagating signal quickly steepens into a shock. This shock propagates along the magnetic field lines, reaching the low solar corona, and is followed by the chromospheric plasma, which consists of the cold and dense jet (spicule). This jet exhibits properties of a contact wave (\citealt{Kuzma2017}) and reaches upto a certain height (typically 4-6 Mm) and then returns to the chromosphere. The mean upflow speed was 20-25 km s$^{-1}$. The comparison between the numerical and observational data always enlightens the path to go further. Therefore, we tried to compare the observed properties of the simulated spicule with the various spicule observations. \cite{Beckers1968,Beckers1972} reported that the average height of spicules can vary from 6500 km to 9500 km. Similarly, \cite{Lipp1957} reported the spicule height range from 7000 km to 13,000 km and \cite{Pasa2009} inferred these numbers from 4200 km to 12,200 km. Observed upward velocity of spicules exhibits variations within the range of 
%as it concerns to the observational reports 
20.0-150.0 km s$^{-1}$ (\citealt{DePon2007}) and 3.0-75.0 km s$^{-1}$ (\citealt{Pasa2009}). The numerically obtained value of spicule height ($\sim$ 4.9 Mm) and upward velocity ($\sim$ 20-25 km s$^{-1}$) of the simulated spicule lies in the observed range of height and velocity (\citealt{Beckers1968,Beckers1972,DePon2007,Pasa2009}). It should also be noted that the numerical values of the spicule match with those obtained for MHD (\citealt{Kuzma2017}). % and they \textbf{also match with the observationally reported values of spicules}. 

%showed that the spicule height can be vary from 4200 km to 12200 km. In the present simulation, the ion as well as neutral spicule reaches approximately 4800 km. It is not a trival job to estimate the density and temperature of spicules, therefore, there are very observational works }.

Novelty of our numerical simulations lies in the usage of the two-fluid model, which was not investigated so far. The two-fluid simulations invoke real physical conditions of the lower solar atmosphere and plasma processes within various localized jets. It is noteworthy that a significant role of neutrals associated with confined pseudo-shocks in carrying substantial energy and mass into the overlying solar atmosphere was discovered by \cite{Srivastava2017}. The key findings of our simulations can be summarized as follows. (a) The spicule dynamics is slightly different in ions and neutrals; (b) the core of the spicule is dominated by cold neutrals; (c) the neutral spicule is wider compared to the ion spicule; (d) the ionization remains essentially still with height and horizontal distance within the spicule, only the top of the spicule is significantly ionized; and (e) in later moments of time of the spicule evolution higher amplitude of the slow rarefaction wave for the ion spicule results in a different structure than for the neutral spicule. The general scenario of the rise and the downfall of the spicule is the same for ions and neutrals and it is close to that for MHD (\citealt{Kuzma2017}). The dispatches in gas pressure gradients between ions and neutrals is a key factor to understand the minor differences in the dynamics of ions and neutral spicules. Obviously, the magnetic field does not have any direct influence on the dynamics of neutrals, while the motion of ions is affected by the magnetic field directly. The direct influence of the magnetic field on the ions controls the width of the ion spicule. Therefore, the ion spicule is a very well collimated column of ions with its small width. In the absence of the influence of magnetic field, the neutrals get extra freedom to propagate in the horizontal direction; a constraint on them results from ion-neutral collisions. As a result, the neutral spicule is wider compared to the ion spicule. The V-shaped structure near the base of the ion/neutral spicule is an interesting phenomenon of these simulations, which results from down-falling gas. As ions are affected by magnetic field, the well collimated down-falling ion gas results in, which exerts the gas pressure force producing in the decay phase the sharp V-shaped structure near the base of the ion spicule. 
%On the contrary, neutrals are not guided by the magnetic field, therefore, downflow of the neutral gas has more speard horizontally which produce weak and wider V-shaped structure near the base of neutral spicule in the decay phase. 
%which produce the  which is produced due to uneven distribution of down-falling horizontal velocity across the width of spicule.}
%V-shaped structure near the base of spicules is an interesting phenomena of this simulation, which was occured in previous simulations but not reported (....).  

%Down-falling is the important element of this spicule dynamics, which produce the V-shaped structure near the top of ion as well as neutral spicule. In the similar fasion, down-falling of plasma produce the V-shaped structure near the base of spicule. However, this V-shaped structure at the base of neutral spicule is not  } %, that produces the wider neutral spicule. 

Most of observations reveal that temperature and mass density of a spicule remain essentially constant along its height (\citealt{Beckers1968,Beckers1972}; see also review article by \citealt{Sterling2000}). However, our simulated spicule shows a significant variations in its temperature as well as mass density. The temperature of the spicule is constant upto the height of $y=$ 3.5 Mm, while higher up the temperature falls off abruptly. The constant phase of temperature matches the obervational findings. However, there is a sharp increase in the temperature above $y=4.5$ Mm, which may suggest that the thermal evolution of the cold chromospheric plasma can be important as proposed by \cite{DePon2014}. Overall, the mass density of the spicule is decreasing with height, which is qualitatively consistent with the observed mass densities (\citealt{Beckers1968}). However, the top of the spicule shows slightly higher densities. 
%It is apparent that temperature grows with height in the solar atmosphere. So, temperature is low in the upper chromosphere (i.e., the base of the spicule), and temperature is high in the corona (i.e., the top-part of the spicule). 
The high temperature near the top-part of the spicule significantly ionizes the neutrals, which were taken into account in the present scenario. Recently, \cite{DePon2014} reported that spicules undergo thermal evolution using high-resolution observations. Spicules produce more emission near their bases compared to their apexes at chromospheric temperature (i.e., Ca~{\sc ii} H $\&$ K and Mg~{\sc ii} h $\&$ k lines). However, bases of spicules become dark and the top parts of spicules lead to more emissions at the TR temperature (i.e., Si~{\sc iv}; $T=80$ kK). This observational finding predicts that ionization grows with height within the spicule, which is successfully reproduced by our two-fluid numerical simulations of spicules.

In conclusion, our numerical simulations of the spicule performed within the framework of the two-fluid approach successfully mimics the averaged properties of classical spicules. Ion and neutral spicules follow the similar dynamics in terms of rise time and peak altitude. The core of the spicule exhibits an abundance of neutrals, with growing in time rarefaction of ions.

\begin{acknowledgements}
The authors express their thanks to the referee for his/her comments on the earlier version of the draft. This work was financially supported by the project from the National Science Centre, Poland, (NCN) Grant No. 2014/15/B/ST9/00106. The JOANNA code used in this work was developed by Mr. Darek W\'ojcik. These numerical simulations were performed on the LUNAR cluster at Institute of Mathematics of University of M. Curie-Sk\l{}odowska. Visualization of the simulations data was done with the use of IDL (Interactive Data Language) and VisIt software packages.

\end{acknowledgements}

%{\bf Bibliography}
%\\
{\footnotesize
%W\'ojcik, D., 2017, in preparation
%\bibliographystyle{aa} % style aa.bst
\bibliography{listb.bib} % your references Yourfile.bib 

\begin{thebibliography}{}
\expandafter\ifx\csname natexlab\endcsname\relax\def\natexlab#1{#1}\fi
\providecommand{\url}[1]{\href{#1}{#1}}

\bibitem[{Avrett \& Loeser(2008)}]{Avrett2008}
Avrett, E.~H., \& Loeser, R. 2008, ApJS, 175, 228

\bibitem[{{Beck} {et~al.}(2016){Beck}, {Rezaei}, {Puschmann}, \&
  {Fabbian}}]{Beck2016}
{Beck}, C., {Rezaei}, R., {Puschmann}, K.~G., \& {Fabbian}, D. 2016, \solphys,
  291, 2281

\bibitem[{{Beckers}(1968)}]{Beckers1968}
{Beckers}, J.~M. 1968, \solphys, 3, 367

\bibitem[{{Beckers}(1972)}]{Beckers1972}
---. 1972, ARA\&A, 10, 73

\bibitem[{{Cheng}(1992)}]{Cheng1992}
{Cheng}, Q.-Q. 1992, \aap, 266, 537

\bibitem[{{Courant} {et~al.}(1928){Courant}, {Friedrichs}, \&
  {Lewy}}]{Courant1928}
{Courant}, R., {Friedrichs}, K., \& {Lewy}, H. 1928, Mathematische Annalen,
  100, 32

\bibitem[{{Cranmer} \& {Woolsey}(2015)}]{Cranmer2015}
{Cranmer}, S.~R., \& {Woolsey}, L.~N. 2015, ApJ, 812, 71

\bibitem[{{Cranmer} \& {Woolsey}(2016)}]{Cranmer2016}
---. 2016, ApJ, 822, 119

\bibitem[{{De Pontieu} {et~al.}(2004){De Pontieu}, {Erd{\'e}lyi}, \&
  {James}}]{DePon2004}
{De Pontieu}, B., {Erd{\'e}lyi}, R., \& {James}, S.~P. 2004, \nat, 430, 536

\bibitem[{{De Pontieu} {et~al.}(2007{\natexlab{a}}){De Pontieu}, {Hansteen},
  {Rouppe van der Voort}, {van Noort}, \& {Carlsson}}]{DePon2007a}
{De Pontieu}, B., {Hansteen}, V.~H., {Rouppe van der Voort}, L., {van Noort},
  M., \& {Carlsson}, M. 2007{\natexlab{a}}, \apj, astro-ph/0701786

\bibitem[{{De Pontieu} {et~al.}(2009){De Pontieu}, {McIntosh}, {Hansteen}, \&
  {Schrijver}}]{DePon2009}
{De Pontieu}, B., {McIntosh}, S.~W., {Hansteen}, V.~H., \& {Schrijver}, C.~J.
  2009, \apjl, 701, L1

\bibitem[{{De Pontieu} {et~al.}(2007{\natexlab{b}}){De Pontieu}, {McIntosh},
  {Hansteen}, {Carlsson}, {Schrijver}, {Tarbell}, {Title}, {Shine}, {Suematsu},
  {Tsuneta}, {Katsukawa}, {Ichimoto}, {Shimizu}, \& {Nagata}}]{DePon2007}
{De Pontieu}, B., {McIntosh}, S., {Hansteen}, V.~H., {et~al.}
  2007{\natexlab{b}}, \pasj, 59, S655

\bibitem[{{Haerendel}(1992)}]{Hae1992}
{Haerendel}, G. 1992, \nat, 360, 241

\bibitem[{{Hansteen} {et~al.}(2006){Hansteen}, {De Pontieu}, {Rouppe van der
  Voort}, {van Noort}, \& {Carlsson}}]{Hansteen2006}
{Hansteen}, V.~H., {De Pontieu}, B., {Rouppe van der Voort}, L., {van Noort},
  M., \& {Carlsson}, M. 2006, \apjl, 647, L73

\bibitem[{{Heggland} {et~al.}(2007){Heggland}, {De Pontieu}, \&
  {Hansteen}}]{Heggland2007}
{Heggland}, L., {De Pontieu}, B., \& {Hansteen}, V.~H. 2007, \apj, 666, 1277

\bibitem[{{Hirayama}(1992)}]{Hirayama1992}
{Hirayama}, T. 1992, \solphys, 137, 33

\bibitem[{{Hollweg}(1972)}]{Hollweg1972}
{Hollweg}, J.~V. 1972, Cosmic Electrodynamics, 2, 423

\bibitem[{{Hollweg}(1982)}]{Hollweg1982}
---. 1982, \apj, 257, 345

\bibitem[{{Hollweg} {et~al.}(1982){Hollweg}, {Jackson}, \&
  {Galloway}}]{Hollweg1982a}
{Hollweg}, J.~V., {Jackson}, S., \& {Galloway}, D. 1982, \solphys, 75, 35

\bibitem[{{James} {et~al.}(2003){James}, {Erd{\'e}lyi}, \& {De
  Pontieu}}]{James2003}
{James}, S.~P., {Erd{\'e}lyi}, R., \& {De Pontieu}, B. 2003, \aap, 406, 715

\bibitem[{Konkol {et~al.}(2012)Konkol, Murawski, \& Zaqarashvili}]{Konkol2012}
Konkol, P., Murawski, K., \& Zaqarashvili, T.~V. 2012, A\&A, 537, A96

\bibitem[{{Kopp} \& {Kuperus}(1968)}]{Kopp1968}
{Kopp}, R.~A., \& {Kuperus}, M. 1968, \solphys, 4, 212

\bibitem[{{Kudoh} \& {Shibata}(1999)}]{Kudoh1999}
{Kudoh}, T., \& {Shibata}, K. 1999, \apj, 514, 493

\bibitem[{Ku\'zma {et~al.}(2017)Ku\'zma, Murawski, Zaqarashvili, Konkol, \&
  Mignone}]{Kuzma2017}
Ku\'zma, B., Murawski, K., Zaqarashvili, T.~V., Konkol, P., \& Mignone, A.
  2017, A\&A, 597, 133

\bibitem[{Lippincott(1957)}]{Lipp1957}
Lippincott, S.~L. 1957, Smithsonian Contributions to Astrophysics, 2, 15

\bibitem[{{Madjarska} {et~al.}(2011){Madjarska}, {Vanninathan}, \&
  {Doyle}}]{Mad2011}
{Madjarska}, M.~S., {Vanninathan}, K., \& {Doyle}, J.~G. 2011, \aap, 532, L1

\bibitem[{{Matsuno} \& {Hirayama}(1988)}]{Matsuno1988}
{Matsuno}, K., \& {Hirayama}, T. 1988, \solphys, 117, 21

\bibitem[{{McIntosh} \& {De Pontieu}(2009)}]{McIntosh2009}
{McIntosh}, S.~W., \& {De Pontieu}, B. 2009, \apj, 707, 524

\bibitem[{{McIntosh} {et~al.}(2010){McIntosh}, {Innes}, {De Pontieu}, \&
  {Leamon}}]{McIntosh2010}
{McIntosh}, S.~W., {Innes}, D.~E., {De Pontieu}, B., \& {Leamon}, R.~J. 2010,
  \aap, 510, L2

\bibitem[{{Moore} \& {Fung}(1972)}]{Moore1972}
{Moore}, R.~L., \& {Fung}, P.~C.~W. 1972, \solphys, 23, 78

\bibitem[{{Murawski} \& {Zaqarashvili}(2010)}]{Murawski2010}
{Murawski}, K., \& {Zaqarashvili}, T.~V. 2010, \aap, 519, A8

\bibitem[{Nakariakov \& Verwichte(2005)}]{Nakariakov2005}
Nakariakov, V.~M., \& Verwichte, E. 2005, LRSP, 2, 3

\bibitem[{{Nishikawa}(1988)}]{Nisikawa1988}
{Nishikawa}, T. 1988, \pasj, 40, 613

\bibitem[{Pasachoff {et~al.}(2009)Pasachoff, Jacobson, \& Sterling}]{Pasa2009}
Pasachoff, J.~M., Jacobson, W.~A., \& Sterling, A.~C. 2009, SoPh, 260, 59

\bibitem[{{Pereira} {et~al.}(2012){Pereira}, {De Pontieu}, \&
  {Carlsson}}]{Per2012}
{Pereira}, T.~M.~D., {De Pontieu}, B., \& {Carlsson}, M. 2012, \apj, 759, 18

\bibitem[{{Pereira} {et~al.}(2016){Pereira}, {Rouppe van der Voort}, \&
  {Carlsson}}]{Per2016}
{Pereira}, T.~M.~D., {Rouppe van der Voort}, L., \& {Carlsson}, M. 2016, \apj,
  824, 65

\bibitem[{{Pereira} {et~al.}(2014){Pereira}, {De Pontieu}, {Carlsson},
  {Hansteen}, {Tarbell}, {Lemen}, {Title}, {Boerner}, {Hurlburt}, {W{\"u}lser},
  {Mart{\'{\i}}nez-Sykora}, {Kleint}, {Golub}, {McKillop}, {Reeves}, {Saar},
  {Testa}, {Tian}, {Jaeggli}, \& {Kankelborg}}]{DePon2014}
{Pereira}, T.~M.~D., {De Pontieu}, B., {Carlsson}, M., {et~al.} 2014, \apjl,
  792, L15

\bibitem[{{Roberts}(1979)}]{Roberts1979}
{Roberts}, B. 1979, \solphys, 61, 23

\bibitem[{{Roberts}(1945)}]{Roberts1945}
{Roberts}, W.~O. 1945, \apj, 101, 136

\bibitem[{{Rouppe van der Voort} {et~al.}(2015){Rouppe van der Voort}, {De
  Pontieu}, {Pereira}, {Carlsson}, \& {Hansteen}}]{Rouppe2015}
{Rouppe van der Voort}, L., {De Pontieu}, B., {Pereira}, T.~M.~D., {Carlsson},
  M., \& {Hansteen}, V. 2015, \apjl, 799, L3

\bibitem[{{Secchi}(1887)}]{Secchi1877}
{Secchi}, P.~A. 1887, Le Soleil, vol. 2 (Gauthier-villas, Paris)

\bibitem[{{Skogsrud} {et~al.}(2014){Skogsrud}, {Rouppe van der Voort}, \& {De
  Pontieu}}]{Skog2014}
{Skogsrud}, H., {Rouppe van der Voort}, L., \& {De Pontieu}, B. 2014, \apjl,
  795, L23

\bibitem[{{Skogsrud} {et~al.}(2015){Skogsrud}, {Rouppe van der Voort}, {De
  Pontieu}, \& {Pereira}}]{Skog2015}
{Skogsrud}, H., {Rouppe van der Voort}, L., {De Pontieu}, B., \& {Pereira},
  T.~M.~D. 2015, \apj, 806, 170

\bibitem[{{Smith} \& {Sakai}(2008)}]{Sakai2008}
{Smith}, P.~D., \& {Sakai}, J.~I. 2008, A\&A, 486, 569

\bibitem[{{Srivastava} {et~al.}(2017){Srivastava}, Murawski, Ku\'zma,
  Zaqarashvili, Stangalini, Musielak, W\'ojcik, Doyle, \&
  Dwivedi}]{Srivastava2017}
{Srivastava}, A.~K., Murawski, K., Ku\'zma, B., {et~al.} 2017, NatAs, under
  review

\bibitem[{{Sterling}(2000)}]{Sterling2000}
{Sterling}, A.~C. 2000, \solphys, 196, 79

\bibitem[{{Sterling} \& {Hollweg}(1984)}]{Sterling1984}
{Sterling}, A.~C., \& {Hollweg}, J.~V. 1984, ApJ, 285, 843

\bibitem[{{Sterling} \& {Mariska}(1990)}]{Sterling1990}
{Sterling}, A.~C., \& {Mariska}, J.~T. 1990, \apj, 349, 647

\bibitem[{{Sterling} \& {Moore}(2016)}]{Sterling2016}
{Sterling}, A.~C., \& {Moore}, R.~L. 2016, ApJL, 828, L9

\bibitem[{{Sterling} {et~al.}(2010){Sterling}, {Moore}, \&
  {DeForest}}]{Sterling2010}
{Sterling}, A.~C., {Moore}, R.~L., \& {DeForest}, C.~E. 2010, \apjl, 714, L1

\bibitem[{{Sterling} {et~al.}(1993){Sterling}, {Shibata}, \&
  {Mariska}}]{Sterling1993}
{Sterling}, A.~C., {Shibata}, K., \& {Mariska}, J.~T. 1993, \apj, 407, 778

\bibitem[{{Suematsu}(1998)}]{Suematsu1998}
{Suematsu}, Y. 1998, in Solar Jets and Coronal Plumes, ESA Special Publication,
  Vol. 421, ed. T.-D. Guyenne, 19

\bibitem[{{Suematsu} {et~al.}(2008){Suematsu}, {Ichimoto}, {Katsukawa},
  {Shimizu}, {Okamoto}, {Tsuneta}, {Tarbell}, \& {Shine}}]{Suematsu2008}
{Suematsu}, Y., {Ichimoto}, K., {Katsukawa}, Y., {et~al.} 2008, in Astronomical
  Society of the Pacific Conference Series, Vol. 397, First Results From
  Hinode, ed. S.~A. {Matthews}, J.~M. {Davis}, \& L.~K. {Harra}, 27

\bibitem[{{Suematsu} {et~al.}(1982){Suematsu}, {Shibata}, {Neshikawa}, \&
  {Kitai}}]{Suematsu1982}
{Suematsu}, Y., {Shibata}, K., {Neshikawa}, T., \& {Kitai}, R. 1982, \solphys,
  75, 99

\bibitem[{{Suematsu} {et~al.}(1995){Suematsu}, {Wang}, \&
  {Zirin}}]{Suematsu1995}
{Suematsu}, Y., {Wang}, H., \& {Zirin}, H. 1995, \apj, 450, 411

\bibitem[{{Tian} {et~al.}(2011){Tian}, {McIntosh}, {Habbal}, \&
  {He}}]{Tian2011}
{Tian}, H., {McIntosh}, S.~W., {Habbal}, S.~R., \& {He}, J. 2011, \apj, 736,
  130

\bibitem[{{Tsiropoula} {et~al.}(2012){Tsiropoula}, {Tziotziou}, {Kontogiannis},
  {Madjarska}, {Doyle}, \& {Suematsu}}]{Tsir2012}
{Tsiropoula}, G., {Tziotziou}, K., {Kontogiannis}, I., {et~al.} 2012, \ssr,
  169, 181

\bibitem[{{W{\'o}jcik} {et~al.}(2017){W{\'o}jcik}, {Murawski}, \&
  Zaqarashvili}]{Woj2017}
{W{\'o}jcik}, D., {Murawski}, K., \& Zaqarashvili, T.~V. 2017, A\&A, in
  preparation

\bibitem[{{Zaqarashvili} \& {Erd{\'e}lyi}(2009)}]{Zaq2009}
{Zaqarashvili}, T.~V., \& {Erd{\'e}lyi}, R. 2009, \ssr, 149, 355

\bibitem[{Zaqarashvili {et~al.}(2011)Zaqarashvili, Khodachenko, \&
  Rucker}]{Zaqarashvili2011}
Zaqarashvili, T.~V., Khodachenko, M.~L., \& Rucker, H.~O. 2011, A\&A, 529, A82

\end{thebibliography}
}
\end{document}